\documentclass{article}
\pdfoutput=1

\usepackage{arxiv}
\usepackage[utf8]{inputenc} 
\usepackage[T1]{fontenc} 
\usepackage{hyperref} 
\hypersetup{
	colorlinks = true,
	linkcolor = black,
	anchorcolor = black,
	citecolor = black,
	filecolor = black,
	urlcolor = blue
}
\usepackage{booktabs} 
\usepackage{amsfonts} 
\usepackage{nicefrac} 
\usepackage{microtype} 
\usepackage{amsmath,bm}
\usepackage{graphicx}
\usepackage[flushleft]{threeparttable}
\usepackage{multirow}
\usepackage{natbib}
\usepackage{makecell}
\usepackage{caption}
\usepackage{subcaption}   
\usepackage{nicematrix}
\usepackage{float}

\newcommand{\figurehere}[1]{\begin{center}%
		=========================\\%
		Insert Figure #1 about here\\%
		=========================\\%
\end{center}}
\newcommand{\tablehere}[1]{\begin{center}%
		=========================\\%
		Insert Table #1 about here\\%
		=========================\\%
\end{center}}

\usepackage{array}
\newcommand{\PreserveBackslash}[1]{\let\temp=\\#1\let\\=\temp}
\newcolumntype{C}[1]{>{\PreserveBackslash\centering}p{#1}}
\newcolumntype{R}[1]{>{\PreserveBackslash\raggedleft}p{#1}}
\newcolumntype{L}[1]{>{\PreserveBackslash\raggedright}p{#1}}

\title{Estimating Rate of Change for nonlinear Trajectories in the Framework of Individual Measurement Occasions: A New Perspective on Growth Curves}

\author{
  Jin Liu \thanks{CONTACT Jin Liu Email: Veronica.Liu0206@gmail.com. \textcircled{c}2023, Behavior Research Methods. This paper is not the copy of record and may not exactly replicate the final, authoritative version of the article. Please do not copy or cite without authors' permission. The final article will be available, upon publication, via its DOI.}\\
Department of Biostatistics\\
Virginia Commonwealth University \\
 \And
Robert A. Perera\\
Department of Biostatistics\\
Virginia Commonwealth University \\
}

\begin{document}

\maketitle
\begin{abstract}
Researchers are often interested in examining between-individual differences in within-individual processes. If the process under investigation is tracked for a long time, its trajectory may show a certain degree of nonlinearity, so that the rate-of-change is not constant. A fundamental goal of modeling such nonlinear processes is to estimate model parameters that reflect meaningful aspects of change, including the parameters related to change and other parameters that shed light on substantive hypotheses. However, if the measurement occasion is unstructured, existing models cannot simultaneously estimate these two types of parameters. This article has three goals. First, we view the change over time as the area under the curve (AUC) of the rate-of-change versus time ($r-t$) graph. Second, using the instantaneous rate-of-change midway through a time interval to approximate the average rate-of-change during that interval, we propose a new specification to describe longitudinal processes. In addition to obtaining the individual change-related parameters and other parameters related to specific research questions, the new specification allows for unequally-space study waves and individual measurement occasions around each wave. Third, we derive the model-based interval-specific change and change-from-baseline, two common measures to evaluate change over time. We evaluate the proposed specification through a simulation study and a real-world data analysis. We also provide \textit{OpenMx} and \textit{Mplus 8} code for each model with the novel specification.
\end{abstract}


\setcounter{secnumdepth}{3}
\section*{Introduction}\label{Intro}
Longitudinal data widely exist in multiple fields, including psychology, education, biomedicine, and behavioral sciences. Analysis of this type of data can provide insights into between-individual differences in within-individual processes. There are multiple perspectives to evaluate within-individual processes: (1) growth status, (2) rate-of-change, (3) change occurs during a time interval and (4) change-from-baseline. Growth status describes the overall trend, while rate-of-change allows for an understanding of the speed of growth. The other two measures, change occurs during a time interval and change from baseline, can be viewed as an accumulative value of rate-of-change during the time interval and the time since baseline, respectively\footnote{An analog in calculus may help understand these four metrics. Suppose we utilize a function to describe growth status; therefore, the rate-of-change can be viewed as the function's first derivative with respect to time $t$. We then obtain the change that occurs during $t_{1}$ to $t_{2}$ by integrating the first derivative from $t_{1}$ to $t_{2}$ and have the change from baseline at $t_{2}$ by integrating the first derivative from $0$ to $t_{2}$.}. The linear trend is the most commonly used function when fitting longitudinal records due to its simplicity and interpretability. Two free coefficients are used to model the growth status of an individual linear trajectory: the intercept and the constant linear slope, representing the outcome of interest at a specific time point (usually at the first study wave) and the rate-of-change over the study duration, respectively. The intercept and the slope are allowed to vary from person to person in commonly used longitudinal modeling frameworks, such as mixed-effect and latent growth curve models. It is also straightforward to obtain the change that occurs during a time interval and the change from baseline since the rate-of-change of this functional form is constant. Therefore, the investigation of individual differences in the intercept and slope is able to provide sufficient information on linear trajectories \citep{Biesanz2004Linear, Zhang2012LCSM, Grimm2013LCSM}. 

However, if the process under examination is followed for a long time period, the trajectory of the longitudinal process may show a certain degree of nonlinearity over time. It is a challenge to estimate the rate-of-change directly. For example, the rate-of-change of the quadratic function (i.e., $y=a\times t^{2}+b\times t$), a commonly used nonlinear curve, consists of a linear coefficient whose instantaneous rate-of-change is $b$, and a quadratic coefficient, $a$, whose instantaneous rate-of-change is $2\times a\times t$. Multiple existing studies have discussed the rate-of-change of nonlinear trajectory. For example, \citet{Kelley2008ARC} delineated the average rate-of-change (ARC) as the change in the outcome measurement divided by the change in time during a specific time interval and demonstrated that the ARC and the slope from the straight-line change model (SLCM) are not equal to each other in general. By calculating the bias and discrepancy factor between the ARC and the SLCM, this work demonstrated that it is problematic to employ the mean slope from SLCM to estimate the mean ARC across individuals. In the empirical application, the authors demonstrated that all three methods lead to biased estimates with over $30\%$ underestimation for a logistic growth curve at best. In addition, \citet{Kelley2009ARC} extended such descriptions and demonstrations to the continuous-time models. Although these earlier studies have successfully shown the conditions to obtain an unbiased estimate of ARC from SLCM, there is no recommendation for the optimal way to estimate the ARC. Therefore, it is also a challenge to derive an accumulated ARC value over time (i.e., the change that occurs in a time interval or change from baseline).

Another challenge of longitudinal data analysis is related to the measurement occasion. First, researchers may not record a longitudinal outcome at a constant frequency, resulting in unequal intervals between study waves. For example, in some popular longitudinal datasets, such as the Early Childhood Longitudinal Research Project, assessments are collected more frequently in the early stage of child development \citep{Le2011ECLS}. Further, the measurement times of each research wave \citep{Finkel2003cognitive, Mehta2000people} may be different. If the time is measured accurately, unstructured measurement occasions will appear. For example, if we evaluate students' academic performance for each grade/semester, we have a regular measurement schedule, but if we evaluate their academic performance based on their actual age, the measurement time is individually different \cite [Chapter~4] {Grimm2016growth}. Researchers have performed simulation studies and shown that the neglect of individual differences in measurement occasions may lead to inadmissible estimates, such as overestimated intra-individual variation and underestimated inter-individual differences \citep{Blozis2008coding, Coulombe2015ignoring}.

Multiple longitudinal modeling frameworks have been proposed and developed to address the above challenges. Some frameworks, such as mixed-effects (or multi-level) models \citep{Harville1977linearmixed, Lindstrom1990nonlinearmixed, Laird1982random, hedeker2006longitudinal, Pinheiro1994mixed, Vonesh1992mixed, Bryk1987hierarchical} and latent growth curve models (LGCMs) \citep{Grimm2016growth, Ram2007LGC, Duncan2000behavior, Duncan2013latent, Bollen2005LGC}, are usually employed to model trajectories, while some other frameworks, such as latent change score models (LCSMs), are used to model rate-of-change \citep{Zhang2012LCSM, Grimm2013LCM1, Grimm2013LCM2}. Each framework is useful in addressing some of the above challenges, but not all. For example, the LGCM, which is mathematically equivalent to the mixed-effects model in the majority of cases \citep{Bauer2003Estimating, Curran2003Multilevel}, is capable of modeling linear or nonlinear trajectories and allows for unequally-spaced study waves and individual measurement occasions\footnote{The time structure with unequal intervals and individual measurement occasions is also referred to as `continuous time' \citep{Driver2017ct, Driver2018ct}. One difference between the models discussed in the article and the `continuous-time' models is that the former can estimate growth parameters related to developmental theory, making it easier to formulate hypotheses, while the latter is used to analyze dynamic processes. So we do not use the term `continuous-time' to avoid confusion.} around each wave by using the definition variable approach \citep{Sterba2014individually, Preacher2015repara, Liu2019BLSGM}. The `definition variable' is defined as an observed variable that adjusts model coefficients to individual-specific values \citep{Mehta2000people, Mehta2005people}. In the LGCM, these individual-specific values are individually different time points. Details of the specification and estimation of the LGCM with individual measurement times are available in earlier studies such as \citet{Sterba2014individually, Preacher2015repara, Liu2019BLSGM}. However, the LGCM cannot estimate the rate-of-change without reparameterization \citep{Preacher2015repara}, except for the linear one \citep{Zhang2012LCSM}, or more broadly, the model with a single between-individual coefficient affecting time \citep{Grimm2013LCM1}. \citet{Kelley2008ARC} and \citet{Kelley2009ARC} pointed out that one may derive ARC from predicted scores (i.e., predicted growth status) by fitting a LGCM with the `correct functional form'. However, it is typically not possible to know the `correct functional form' in practice since it usually does not exist in a real-world scenario. Instead, only an optimal functional form could be determined, often depending on trajectory shapes and, more importantly, specific research questions of interest. Although recent studies, for example, \citet{Preacher2015repara} have utilized this proposal and developed a reparameterized method to estimate fixed and random effects of ARCs, to our knowledge, no theoretical work or simulation studies have been performed to prove this proposal.

On the contrary, the LCSM can be used to estimate the individual rate-of-change by taking the first-order derivative of the corresponding LGCM with respect to time \textit{t} \citep{Zhang2012LCSM, Grimm2013LCM1, Grimm2013LCM2}. The LCSM allows for an explicit estimation of the mean and variance of the rate-of-change over time and then a direct examination of the individual differences in the rate-of-change and its relationship with covariates. However, in its original version, the LCSM assumes that time is discrete; therefore, the measurements are equally spaced, and each individual needs to be measured or assumed to be measured at the same time points. Existing studies have proposed to solve these challenges by involving a constant time period at the latent variable level. For example, \citet{McArdle2001LCSM} proposed adding phantom variables to keep equally-spaced time intervals. \citet[Chapter~18]{Grimm2016growth} demonstrated the method with the Berkeley Growth Study, where participants were followed for $36$ months with records at month $1$, $3$, $6$, $9$, $12$, $15$, $18$, $24$ and $36$. The authors specify a latent true score for each month during the three years by adding a phantom variable for each month without records. This method successfully solved the challenge of unequally-spaced measurement occasions. However, including phantom variables leads to a complex model specification, especially when records are unavailable for most time units. A recent study, \citet{Grimm2018Individually}, has demonstrated that the time can also be continuous and constructed the LCSMs with individual measurement occasions with the \textit{NLMIXED} procedure in \textit{SAS} or using Bayesian modeling tools such as \textit{JAGS} or \textit{WinBUGS}. Specifically, similar to the idea of phantom variables, the authors proposed specifying a latent true score for each point in time between the minimum and maximum values based on the selected time scale. For example, to analyze longitudinal math scores from NLSY-CYA with seven waves of records (from Grade $2$ to Grade $8$), they had to specify $94$ latent true scores since the minimum and maximum age-in-months were $82$ and $175$. The model specification will be more complicated if the time variable is measured with decimal places. For the above example, suppose age-in-month ranges between $82.01$ and $175.01$. The number of potential latent true scores would increase $100$-fold since the specification only allows for 1-step moves for the time metric. The authors propose using the \textit{NLMIXED} procedure in \textit{SAS} or Bayesian tools to allow for adding a loop when specifying a model, which simplifies the model specification. However, it is not straightforward to estimate growth coefficients at the individual level or a derived parameter (i.e., accumulative values in change) as one can do with the structural equation modeling (SEM) software, such as \textit{Mplus 8} and \textit{OpenMx}. The authors demonstrated the proposed method for the dual change score model and stated the limitation of this method in addition to those stated above. For example, the Bayesian tools require intensive computational resources\footnote{In an example provided in \citet{Grimm2018Individually}, a JAGS model converged for all parameters after 50,000 samples and took over two hours.}. Moreover, these existing methods may also produce biased estimates in growth coefficients and then biased rate-of-change when being extended to analyze LCSM with a nonlinear parametric functional form as demonstrated by a simulation study \citep{Liu2022LCSM_JB}. We will also elaborate on it in the following sections.

In this study, we propose a novel specification for LCSMs to directly estimate the mean and variance of the rate-of-change for nonparametric and nonlinear parametric growth trajectories in the framework of individual measurement occasions. Specifically, we view the growth over time as the area under the curve (AUC) of the rate-of-change versus time (\textit{r-t}) graph and propose a novel specification to model rate-of-change and accumulative values in change. More importantly, the individual measurement occasions are modeled through `definition variable' approaches, the same as those in the LGCMs. We demonstrate the proposed specification for one nonparametric functional form (i.e., piecewise linear curve) and three commonly used parametric functions, including quadratic, negative exponential, and Jenss-Bayley curves. The new specification aims to provide a more accurate estimate of growth coefficients and rate-of-change than the existing LCSM specification computationally efficiently. Additionally, with the novel specification, it is easier to estimate accumulative values in change, such as the change that occurs in a time interval and change-from-baseline, at the individual level. We briefly present LGCMs and LCSMs with the nonparametric and three parametric functional forms below. 

\subsection*{Introduction of Latent Growth Curve Modeling Framework}\label{I:LGCMs}
The LGCM is a modeling framework in the SEM family, focusing on analyzing growth status. This section briefly introduces this modeling framework with an overview of commonly used functions, including quadratic, exponential, Jenss-Bayley, and nonparametric functions. In the SEM terminology, a typical growth curve model is fit as a common factor model with a mean structure \citep{Tucker1958functional, Rao1958growth, Meredith1990latent}, and the factors in a LGCM are often called growth factors since they determine the shape of growth curves. A LGCM can be expressed as $\boldsymbol{y}_{i}=\boldsymbol{\Lambda}\times\boldsymbol{\eta}_{i}+\boldsymbol{\epsilon}_{i}$, where $\boldsymbol{y}_{i}$ is a $J\times1$ vector of the repeated measures of the $i^{th}$ individual (in which $J$ is the number of measurements), $\boldsymbol{\eta}_{i}$ is a $k\times1$ vector of latent growth factors for individual $i$ (where $k$ is the number of growth factors), and $\boldsymbol{\Lambda}$ is a $J\times k$ matrix of the corresponding factor loadings. Additionally, $\boldsymbol{\epsilon}_{i}$ is a $J\times1$ vector of residuals of individual $i$. Table \ref{tbl:model_summary} provides the LGCM with several commonly used parametric functions and a nonparametric function, and the interpretation of coefficients related to developmental theory for each functional form. In addition, we plot the growth curve for each function in Figure \ref{fig:growth}.

\tablehere{1}

\figurehere{1}

One advantage of a parametric LGCM is that its parameters are potentially related to theory, allowing researchers to formulate hypotheses more easily. For example, the quadratic component of the change in the quadratic function is related to the change in rate-of-change (i.e., the acceleration). In addition, researchers often view the asymptotic level in the negative exponential function as an individual limit to reflect the individual's capacity. However, we may need more coefficients (e.g., a higher degree of polynomial functions) or nonlinear growth factors (e.g., an individual level of coefficient $b$ or $c$ in Table \ref{tbl:model_summary}) to describe more complex nonlinear change patterns, which may lead to a non-parsimonious model or a model that involves an approximation process. 

Alternatively, we can employ nonparametric functional forms. One example of a nonparametric LGCM is a latent basis growth model (LBGM) \citep{Grimm2011LBGM}, or a shape-factor or free-loading model \citep{Bollen2006LGC, McArdle1986LBGM}. It is a versatile tool for exploring nonlinear trajectories since it does not require any function to describe the change patterns. In the LBGM, there are two growth factors (i.e., $k=2$), a factor indicating the initial status ($\eta_{0i}$) and a shape factor ($\eta_{1i}$). For model identification considerations, we need to fix the intercept factor loadings and any two loadings from the shape factor\footnote{The loading from the shape factor to the first measurement is zero since the intercept is sufficient to indicate the initial status, and thus, we only need to fix one loading among others.}. There are multiple ways to scale and specify the shape factor of a LBGM. For example, if we scale the shape factor as the change from baseline to the first post-baseline time, the unknown loading, $\lambda_{j}$ ($j=2,3,\ldots, J-1$), represents the quotient of the change-from-baseline at the $j^{th}$ post-baseline time to the shape factor. Earlier studies, such as \citet{Sterba2014individually}, have documented how to fit LGCMs with nonlinear parametric and nonparametric functions with individual measurement occasions through `definition variables' approach\footnote{\citet{Sterba2014individually} allows for individual measurement occasions for parametric functions by specifying individual-specific time points. For the LBGM, with two growth factors, the intercept and shape factor scaled as the average net change per time unit, the author expressed the corresponding shape factor loadings as the sum of individual measurement time and departures from linearity. Refer to \citet{Sterba2014individually} for more technical details.}.

\subsection*{Introduction of Latent Change Score Modeling Framework}\label{I:LCSMs}
As discussed above, while LGCMs are available with multiple parametric and nonparametric functions (e.g., the latent basis growth model) and can be fit in the framework of individual measurement occasions, their focus is to characterize the time-dependent growth status. Therefore, from nonlinear LGCMs, we cannot estimate the rate-of-change at the individual level, which is one primary research interest for longitudinal processes, nor the cumulative value of the rate-of-change. In this section, we introduce another modeling framework, LCSMs, which emphasizes time-dependent change with an overview of the four functional forms as above. 

LCSMs, which are also referred to as latent difference score models \citep{McArdle2001LCM1, McArdle2001LCM2, McArdle2009LCM}, were developed to integrate difference equations into the SEM framework. In the LCSM, the difference scores are sequential temporal states of a longitudinal outcome. Specification of the LCSM starts from the idea of classical test theory: for each individual, the observed score at a specific occasion can be decomposed into a latent true score and a residual\footnote{In classical test theory, the difference between the observed and true scores is usually referred to as a measurement error. However, $\epsilon_{ij}$ is usually called a residual or unique score at time $j$ of individual $i$ in the LGCM and LCSM literature. This manuscript then follows the convention in the LGCM and LCSM literature.}
\begin{equation}\nonumber
y_{ij}=y^{\ast}_{ij}+\epsilon_{ij},
\end{equation}
in which $y_{ij}$, $y^{\ast}_{ij}$ and $\epsilon_{ij}$ are the observed score, the latent true score, and the residual of individual $i$ at time $j$, respectively. The true score at time $j$ (i.e., $y^{\ast}_{ij}$) can be further expressed as a linear combination of the true score at the prior time point $j-1$ (i.e., $y^{\ast}_{i(j-1)}$) and the latent change score from time $j-1$ to time $j$ (i.e., $\delta y_{ij}$)
\begin{equation}\nonumber
y^{\ast}_{ij}=y^{\ast}_{i(j-1)}+\delta y_{ij}.
\end{equation}
The parameters that can be estimated from a LCSM directly include (1) the mean and variance of the initial status, (2) the mean and variance of each latent change score, and (3) the residual variance. The estimated means and variances of the change scores allow for examining the within-individual changes in between-individual differences in the rate-of-change. 

\citet{McArdle2001LCM1} and \citet[Chapter~11]{Grimm2016growth} have shown that the LGCM with the nonparametric function (i.e., the LBGM) can also be fit in the LCSM framework. They expressed a linear latent curve model in the LCSM framework in which the individual latent change score is a constant and scaled the score by time-varying basis coefficients. Therefore, a latent change score can be written as 
\begin{equation}\label{eq:LBGM0}
\delta y_{ij}=a_{j}\times\eta_{1i},
\end{equation}
where $a_{j}$ is the time-varying basis coefficient of the constant factor $\eta_{1i}$ at time $j$. Similar to the LBGM in the LGCM framework, we can fix the basis coefficient of the first post-baseline period to $1$ for model identification, so $\eta_{1i}$ represents the slope of this interval. In addition, \citet{Zhang2012LCSM} and \citet{Grimm2013LCSM} have shown that a LGCM with a parametric function also has its corresponding LCSM. We provide the equation and plot of the rate-of-change for the nonparametric and each parametric LCSM in Table \ref{tbl:model_summary} and Figure \ref{fig:rate}, respectively. When estimating these parametric nonlinear LCSMs, the rate at $t_{j}$ is employed to approximate the rate during the interval ($t_{j-1}$, $t_{j}$) in the existing specification. The coefficients in each parametric LCSM have the same interpretation as those in the corresponding LGCM. In addition to the parameters that contribute to the rate-of-change, we also need to estimate the mean and variance of the initial status when fitting a parametric LCSM.

\figurehere{2}

With these nonlinear LCSMs, we can simultaneously estimate individuals' rate-of-change over time and the parameters related to developmental theory. Yet there are multiple assumptions for the model specification of the existing LCSM framework. First, it is usually assumed that the measurement times are equally-spaced. For example, in Equation \ref{eq:LBGM0}, $\delta y_{ij}$ is the change from $t=j-1$ to $t=j$, and $a_{j}\times\eta_{1i}$ is the slope of the time period between $t=j-1$ and $t=j$. They are mathematically equivalent if the slope is constant within the time interval and the interval between two measurement occasions is scaled. The constant slope assumption is established for the LBGM\footnote{When defining a LBGM, it is reasonable to assume that the rate-of-change in each time interval between two consecutive measurement occasions is constant for model identification. Therefore, the latent basis growth curve with $J$ measurement occasions can be viewed as a linear piecewise function with $J-1$ segments.} (see Figure \ref{fig:rate_nonp}), yet it is not satisfied in any parametric nonlinear LCSM (see Figures \ref{fig:rate_quad}-\ref{fig:rate_JB}). Moreover, the assumption to ensure scaled time intervals is that the measurement times are equally-spaced. However, it is only sometimes valid in a real-world scenario; researchers often tend to record more frequently in the early stages of longitudinal studies, where changes are usually more rapid.

We use the rate-of-change versus time ($r-t$) graphs in Figure \ref{fig:rate} to illustrate our point. According to the fundamental theorem of calculus, the AUC in a time interval of the $r-t$ graph is the amount of change in that interval. As shown in Figure \ref{fig:rate_nonp}, the change score from $t=0$ to $t=1$ is $5$, which is numerically equal to the slope of this period since the interval is scaled. However, this equivalence does not hold for the score from $t=2$ to $t=4$. We have a similar challenge for parametric nonlinear LCSMs. Some earlier studies, such as \citet{McArdle2001LCSM} and \citet[Chapter~18]{Grimm2016growth}, successfully solved the challenge of unequally-spaced measurement occasions by adding phantom variables and specifying a latent change score for each scaled period in the study duration. Suppose that we skip the measurements at $t=3$. Using this method, we can define the latent change score from $t=2$ to $t=4$ as the sum of the change from $t=2$ to $t=3$ and that from $t=3$ to $t=4$. However, the model specification becomes complicated if the study duration is long or the scaled time interval is short. 

As demonstrated by a simulation study in \citet{Liu2022LCSM_JB}, this approach may yield biased estimates when being extended to fit a LCSM with a parametric functional form where the slope is not constant in each interval ($t_{j-1}, t_{j}$). In the existing framework demonstrated by \citet[Chapter~18]{Grimm2016growth}, the rate-of-change at $t_{j}$ is utilized to approximate ARC during the interval ($t_{j-1}, t_{j}$), which results in bias: the change score is underestimated when the rate-of-change decreases. We illustrate this point with the grey boxes in Figure \ref{fig:rate_exp}. With the method proposed by \citet[Chapter~18]{Grimm2016growth}, the approximated change score from $t=2$ to $t=3$ and from $t=4$ to $t=6$ are enclosed by the solid and dashed boxes, respectively. Both are smaller than the true AUC of the corresponding time interval, leading to underestimated change scores. Similarly, the change score is overrated if the rate-of-change increases. Conceptually, one may still have the issue of biased estimates if extending \citet{Grimm2018Individually} to fit a LCSM with a parametric functional form, since the method with individually varying time points proposed by this work can be viewed as an extension of the idea of phantom variables with smaller time units, depending on the minimum and maximum values and the decimal place of the recorded times.

To solve these challenges, we propose a novel specification for LCSMs. Specifically, we define the change score in an interval as the AUC of that interval, which can be further expressed as the product of the interval-specific ARC and the interval length. Note that the ARCs are accurate values for the nonparametric functional form since each interval-specific ARC is constant, as shown in Figure \ref{fig:rate_nonp}. On the contrary, we employ the instantaneous rate-of-change midway through an interval to approximate the interval-specific ARC for the three parametric functional forms. This novel specification aims to provide more accurate estimates and allow for the extension of the definition variable approach to fit the LCSM in the framework of individual measurement occasions, which we will further discuss in the Method section. We also aim to estimate derived parameters such as the change that occurs in a time interval or the change from baseline since we fit the proposed models in SEM software \textit{OpenMx}. Additionally, we are interested in obtaining factor scores for each growth factor, rate-of-change, change in each interval, and change-from-baseline, which allow for an evaluation of the longitudinal process of each individual.

The rest of this article is organized as follows. First, we describe the model specification and estimation of each extended LCSM with the above four functions. In the following section, we show the design of a Monte Carlo simulation to evaluate the novel specification. Specifically, we present performance metrics, including the relative bias, the empirical standard error (SE), the relative root-mean-squared-error (RMSE), and the empirical coverage probability (CP) for a nominal $95\%$ confidence interval of parameters of interest. We also compare each LCSM with the novel specification with the corresponding LGCM. In the Application section, we analyze a real-world dataset to demonstrate how to fit and interpret the LGCM and LCSM. Finally, we discuss practical considerations, methodological considerations, and future directions.

\section*{Method}\label{Method}
\subsection*{Model Specification of Nonparametric Latent Change Score Models}\label{M:nonpara_specify}
In this section, we present a new specification for the LBGM in the LCSM framework. Following \citet{McArdle2001LCM1} and \citet[Chapter~11]{Grimm2016growth}, we view the LBGM with $J$ measures as a linear piecewise function with $J-1$ segments. For the $i^{th}$ individual, we specify the model as
\begin{align}
&y_{ij}=y^{\ast}_{ij}+\epsilon_{ij},\label{eq:LBGM1}\\
&y^{\ast}_{ij}=\begin{cases}
\eta_{0i}, & \text{if $j=1$}\\
y^{\ast}_{i(j-1)}+dy_{ij}\times(t_{ij}-t_{i(j-1)}), & \text{if $j=2, \dots, J$}
\end{cases},\label{eq:LBGM2}\\
&dy_{ij}=\eta_{1i}\times\gamma_{j-1}\qquad (j=2, \dots, J). \label{eq:LBGM3}
\end{align}
Equations \ref{eq:LBGM1} and \ref{eq:LBGM2} together define the basic setting of a LCSM, where $y_{ij}$, $y^{\ast}_{ij}$, and $\epsilon_{ij}$ are the observed measurement, latent true score, and residual of the $i^{th}$ individual at time $j$, respectively. At baseline (i.e., $j=1$), the true score is the growth factor indicating the initial status ($\eta_{0i}$); at each post-baseline time point (i.e., $j\ge2$), the true score at time $j$ is a linear combination of the score at the prior time point $j-1$ and the amount of true change from time $j-1$ to $j$, which can be further expressed as the product of the time interval ($t_{ij}-t_{i(j-1)}$) and the interval-specific slope ($dy_{ij}$). As shown in Figure \ref{fig:rate_nonp}, this product is the interval-specific AUC of the $r-t$ graph. Note that each time interval is not necessarily equal. The subscript $i$ of $t$ indicates that the measurement times are allowed to be individually different, and so are the time intervals. Note that such individual intervals are the definition variables in the proposed model specification. We scale the shape factor ($\eta_{1i}$) to the slope in the first time intervals. Then each interval specified slope ($dy_{ij}$) can be expressed as the product of $\eta_{1i}$ and the corresponding relative rate $\gamma_{j-1}$, which is also an unknown parameter when $j>2$ in the specified model, as Equation \ref{eq:LBGM3}. Note that one underlying assumption of the above model specification is that the residuals are time-independent. We provide a path diagram of the LBGM with six measurements using the novel specification in Figure \ref{fig:path_nonp}, where we use the diamond shape to denote the definition variables by following \citet{Mehta2005people, Sterba2014individually}, and to illustrate the heterogeneity of the time intervals.

\figurehere{3}

The model defined in Equations \ref{eq:LBGM1}-\ref{eq:LBGM3} can also be expressed in a matrix form as
\begin{equation}\label{eq:matrix_spec1}
\boldsymbol{y}_{i}=\boldsymbol{\Lambda}_{i}\times\boldsymbol{\eta}_{i}+\boldsymbol{\epsilon}_{i},
\end{equation}
where $\boldsymbol{y}_{i}$ is a $J\times1$ vector of the repeated measurements of the $i^{th}$ individual (in which $J$ is the number of measures), $\boldsymbol{\eta}_{i}$ is a $2\times1$ vector of growth factors of which the first element is the initial status and the second element is the slope in the first time interval, and $\boldsymbol{\Lambda}_{i}$ is a $J\times2$ matrix of the corresponding factor loadings,
\begin{equation}\label{eq:LBGM_load}
\boldsymbol{\Lambda}_{i}=\begin{pNiceMatrix}
1 & \Block[fill=gray!36,rounded-corners]{5-1}{}0 \\
1 & \gamma_{1}\times(t_{i2}-t_{i1}) \\
1 & \sum_{j=2}^{3}\gamma_{j-1}\times(t_{ij}-t_{i(j-1)}) \\
\dots & \dots \\
1 & \sum_{j=2}^{J}\gamma_{j-1}\times(t_{ij}-t_{i(j-1)}) \\
\end{pNiceMatrix}.
\end{equation}
The subscript $i$ in $\boldsymbol{\Lambda}_{i}$ indicates that the model is built in the framework of individual measurement occasions. Similar to LGCMs, the first column of $\boldsymbol{\Lambda}_{i}$ is the factor loadings of the intercept, so all loadings are $1$. The $j^{th}$ element in the second column is the cumulative value of the relative rate\footnote{Note that the relative rate from $t_{i1}$ to $t_{i2}$ is fixed as $1$ (i.e., $\gamma_{1}=1$) for identification consideration.} over time up to time $j$, so the product of it and $\eta_{1i}$ represents the change from the initial status, which is also the value of AUC of $r-t$ graph from the start to time $j$. Additionally, $\boldsymbol{\epsilon}_{i}$ is a $J\times1$ vector of residuals of the $i^{th}$ individual. The growth factors $\boldsymbol{\eta}_{i}$ can be further expressed as 
\begin{equation}\label{eq:matrix_spec2}
\boldsymbol{\eta}_{i}=\boldsymbol{\mu}_{\boldsymbol{\eta}}+\boldsymbol{\zeta}_{i},
\end{equation}
in which $\boldsymbol{\mu}_{\boldsymbol{\eta}}$ is the mean vector of the growth factors, and $\boldsymbol{\zeta}_{i}$ is the vector of deviations of individual $i$ from the corresponding mean values of growth factors.

\subsection*{Model Specification of Parametric Latent Change Score Models}\label{M:para_specify}
This section presents the new specification for nonlinear parametric LCSMs. The specification of a parametric model is not as straightforward as that of the nonparametric model since the rate-of-change in each time interval is not constant, as shown in Figures \ref{fig:rate_quad}-\ref{fig:rate_JB}. To solve this challenge, we utilize the instantaneous slope midway through the time interval ($t_{j-1}$, $t_{j}$) to approximate the ARC in that interval similar to \citet{Liu2022LCSM_JB}. We illustrate this idea using grey boxes in Figure \ref{fig:rate_JB}. For example, for the change score from $t=2$ to $t=3$, we use the instantaneous slope at $t=2.5$ to approximate the ARC; therefore, the approximated latent change score is the area in the solid box. Similarly, the approximated change from $t=4$ to $t=6$ is the area in the dashed box, with the ARC approximated by the instantaneous slope at $t=5$. Therefore, each interval-specific AUC for each parametric model is approximated as the product of the interval length and the instantaneous slope midway through the interval.

When specifying the parametric LCSMs, we still need Equation \ref{eq:LBGM1}. We approximate the latent change score of each interval as the product of the instantaneous slope in the middle of the interval and the corresponding interval length and express the latent true score as
\begin{equation}\label{eq:LCSM}
y^{\ast}_{ij}=\begin{cases}
\eta_{0i}, & \text{if $j=1$}\\
y^{\ast}_{i(j-1)}+dy_{ij\_\text{mid}}\times(t_{ij}-t_{i(j-1)}), & \text{if $j=2, \dots, J$}
\end{cases}
\end{equation}
in which $dy_{ij\_\text{mid}}$ is the slope at the midpoint from $j-1$ to $j$. For the $i^{th}$ individual, the instantaneous slope of quadratic, exponential, and Jenss-Bayley curves can be written as follows
\begin{itemize}
\item{Quadratic Function: \begin{align}
dy_{ij\_\text{mid}}
&=\frac{d}{dt}\big(\eta_{0i}+\eta_{1i}\times t+\eta_{2i}\times{t^{2}}+\epsilon_{ij}\big)|_{t=t_{ij\_\text{mid}}}\nonumber\\
&=\eta_{1i}+2\times\eta_{2i}\times t_{ij\_\text{mid}}\label{eq:quad},
\end{align}} \\
\item{Negative Exponential Function: \begin{align}
dy_{ij\_\text{mid}}
&=\frac{d}{dt}\big(\eta_{0i}+\eta_{1i}\times(1-\exp(-b\times t))+\epsilon_{ij}\big)|_{t=t_{ij\_\text{mid}}}\nonumber\\
&=b\times\eta_{1i}\times\exp(-b\times t_{ij\_\text{mid}})\label{eq:exp},
\end{align}} \\
\item{Jenss-Bayley function: \begin{align}
dy_{ij\_\text{mid}}
&=\frac{d}{dt}\big(\eta_{0i}+\eta_{1i}\times t+\eta_{2i}\times(\exp(c\times t)-1)+\epsilon_{j}\big)|_{t=t_{ij\_\text{mid}}}\nonumber\\
&=\eta_{1i}+c\times\eta_{2i}\times\exp(c\times t_{ij\_\text{mid}})\label{eq:JB}.
\end{align}}
\end{itemize}
With this definition, $dy_{ij\_\text{mid}}\times(t_{ij}-t_{i(j-1)})$ defined in Equation \ref{eq:LCSM} is an approximated value of the AUC of a parametric $r-t$ graph (i.e., an approximated value of the interval-specific change) from time $j-1$ to $j$. Each coefficient in Equations \ref{eq:quad}, \ref{eq:exp}, and \ref{eq:JB} is interpreted as the corresponding element introduced in Table \ref{tbl:model_summary}. Similar to the nonparametric LCSM, these parametric nonlinear LCSMs can be expressed in the matrix form as Equations \ref{eq:matrix_spec1} and \ref{eq:matrix_spec2}. For the quadratic, negative exponential, and Jenss-Bayley models, $\boldsymbol{\eta}_{i}$ is a $3\times1$, $2\times1$, and $3\times1$ vector of growth factors, respectively, and their corresponding factor loading matrices are 
\begin{itemize}
\item{Quadratic Function: \begin{equation}\label{eq:quad_load}
\boldsymbol{\Lambda}_{i}=\begin{pNiceMatrix}
1 & \Block[fill=gray!36,rounded-corners]{5-2}{} 0 & 0\\
1 & (t_{i2}-t_{i1}) & 2\times t_{i2\_\text{mid}}\times(t_{i2}-t_{i1})\\
1 & \sum_{j=2}^{3}(t_{ij}-t_{i(j-1)}) & 2\times\sum_{j=2}^{3}t_{ij\_\text{mid}}\times(t_{ij}-t_{i(j-1)})\\
\dots & \dots & \dots \\
1 & \sum_{j=2}^{J}(t_{ij}-t_{i(j-1)}) & 2\times\sum_{j=2}^{J}t_{ij\_\text{mid}}\times(t_{ij}-t_{i(j-1)})\\
\end{pNiceMatrix},
\end{equation}} \\
\item{Negative Exponential Function:  \begin{equation}\label{eq:exp_load}
\boldsymbol{\Lambda}_{i}=\begin{pNiceMatrix}
1 & \Block[fill=gray!36,rounded-corners]{5-1}{}0 \\
1 & b\times\exp(-b\times t_{i2\_\text{mid}})\times(t_{i2}-t_{i1}) \\
1 & b\times\sum_{j=2}^{3}\exp(-b\times t_{ij\_\text{mid}})\times(t_{ij}-t_{i(j-1)}) \\
\dots & \dots \\
1 & b\times\sum_{j=2}^{J}\exp(-b\times t_{ij\_\text{mid}})\times(t_{ij}-t_{i(j-1)}) \\
\end{pNiceMatrix},
\end{equation}} \\
\item{Jenss-Bayley function: \begin{equation}\label{eq:JB_load}
\boldsymbol{\Lambda}_{i}=\begin{pNiceMatrix}
1 & \Block[fill=gray!36,rounded-corners]{5-2}{} 0 & 0\\
1 & (t_{i2}-t_{i1}) & c\times\exp(c\times t_{i2\_\text{mid}})\times(t_{i2}-t_{i1})\\
1 & \sum_{j=2}^{3}(t_{ij}-t_{i(j-1)}) & c\times\sum_{j=2}^{3}\exp(c\times t_{ij\_\text{mid}})\times(t_{ij}-t_{i(j-1)})\\
\dots & \dots & \dots \\
1 & \sum_{j=2}^{J}(t_{ij}-t_{i(j-1)}) & c\times\sum_{j=2}^{J}\exp(c\times t_{ij\_\text{mid}})\times(t_{ij}-t_{i(j-1)})\\
\end{pNiceMatrix}.
\end{equation}}
\end{itemize}
Similar to the first column of $\boldsymbol{\Lambda}_{i}$ in Equation \ref{eq:LBGM_load}, the first column of $\boldsymbol{\Lambda}_{i}$ in Equations \ref{eq:quad_load}-\ref{eq:JB_load} are the factor loadings of the intercept, but they are for the quadratic, negative exponential, and Jenss-Bayley functions, respectively. In addition, the product of the second and third columns of $\boldsymbol{\Lambda}_{i}$ in Equation \ref{eq:quad_load} and the corresponding growth factor represent the cumulative value of the linear slope (i.e., $\eta_{1i}$) and that of the quadratic slope (i.e., $2\times\eta_{2i}\times t_{ij\_\text{mid}}$) over time, respectively. Similarly, the product of the second column of $\boldsymbol{\Lambda}_{i}$ in Equation \ref{eq:exp_load} and its growth factor is the cumulative value of the negative exponential slope (i.e., $b\times\eta_{1i}\times e^{-b\times t_{ij\_\text{mid}}}$) over time, while the product of the second and third columns of $\boldsymbol{\Lambda}_{i}$ in Equation \ref{eq:JB_load} and the corresponding growth factor are the cumulative value of the linear asymptote slope (i.e., $\eta_{1i}$) and that of the exponential slope (i.e., $c\times\eta_{2i}\times e^{c\times t_{ij\_\text{mid}}}$) over time, respectively. 

With such specifications, the $j^{th}$ element of the product of the grey shaded part of each $\boldsymbol{\Lambda}_{i}$ and the corresponding growth factor(s) is interpreted as the change-from-baseline at time $j$ for the corresponding parametric LCSM. Note that the change-from-baseline at a specific measurement occasion is (an approximate value of) the AUC of $r-t$ graph from the baseline to that occasion. We provide path diagrams for these parametric LCSMs (six measurements) using the novel specification in Figures \ref{fig:path_quad}-\ref{fig:path_JB}. 

\subsection*{Model Estimation}\label{M:Estimate}
We make two assumptions to simplify the estimation. First, we assume that the growth factors are normally distributed; that is, $\boldsymbol{\zeta}_{i}\sim \text{MVN}(\boldsymbol{0}, \boldsymbol{\Psi}_{\boldsymbol{\eta}})$, where $\boldsymbol{\Psi}_{\boldsymbol{\eta}}$ is a $2\times2$,  $3\times3$, $2\times2$, and $3\times3$ variance-covariance matrix for the variance-covariance matrix of the growth factors of nonparametric, quadratic, negative exponential, and Jenss-Bayley LCSMs, respectively. We also assume that residuals are independently and identically normally distributed, that is, for the $i^{th}$ individual, $\boldsymbol{\epsilon}_{i}\sim\text{MVN}(\boldsymbol{0}, \theta_{\epsilon}\boldsymbol{I})$, where $\boldsymbol{I}$ is a $J\times J$ identity matrix. Therefore, for the individual $i$, the expected mean vector and the variance-covariance structure of the repeated measurements $\boldsymbol{y}_{i}$ of a LCSM specified in Equations \ref{eq:matrix_spec1} and \ref{eq:matrix_spec2} are expressed as
\begin{equation}\nonumber
\boldsymbol{\mu}_{i}=\boldsymbol{\Lambda}_{i}\boldsymbol{\mu}_{\boldsymbol{\eta}}
\end{equation}
and
\begin{equation}\nonumber
\boldsymbol{\Sigma}_{i}=\boldsymbol{\Lambda}_{i}\boldsymbol{\Psi}_{\boldsymbol{\eta}}\boldsymbol{\Lambda}_{i}^{T}+\theta_{\epsilon}\boldsymbol{I}.
\end{equation}
The parameters of each LCSM given in Equations \ref{eq:matrix_spec1} and \ref{eq:matrix_spec2} include the mean vector and variance-covariance matrix of the growth factors, and the residual variance. In addition, we need to estimate the relative rate of each time interval for the nonparametric LCSM (i.e., $\gamma_{j}$) if $j\ge2$. We also need to estimate the coefficient $b$ for the negative exponential LCSM and coefficient $c$ for the Jenss-Bayley LCSM. The parameters of each LCSM presented above are detailed below
\begin{itemize}
\item{Nonparametric Function (i.e., LBGM): \begin{align}
\boldsymbol{\Theta}_{\text{LBGM}}&=\{\boldsymbol{\mu}_{\boldsymbol{\eta}}, \boldsymbol{\Psi}_{\boldsymbol{\eta}}, \gamma_{2}, \dots, \gamma_{J-1}, \theta_{\epsilon}\}\nonumber\\
&=\{\mu_{\eta_{0}}, \mu_{\eta_{1}}, \psi_{00}, \psi_{01}, \psi_{11}, \gamma_{2}, \dots, \gamma_{J-1}, \theta_{\epsilon}\}\nonumber
\end{align}} \\
\item{Quadratic Function: \begin{align}
\boldsymbol{\Theta}_{\text{QUAD}}&=\{\boldsymbol{\mu}_{\boldsymbol{\eta}}, \boldsymbol{\Psi}_{\boldsymbol{\eta}}, \theta_{\epsilon}\}\nonumber\\
&=\{\mu_{\eta_{0}}, \mu_{\eta_{1}}, \mu_{\eta_{2}}, \psi_{00}, \psi_{01}, \psi_{02}, \psi_{11}, \psi_{12}, \psi_{22}, \theta_{\epsilon}\}\nonumber
\end{align}} \\
\item{Negative Exponential Function: \begin{align}
\boldsymbol{\Theta}_{\text{EXP}}&=\{\boldsymbol{\mu}_{\boldsymbol{\eta}}, \boldsymbol{\Psi}_{\boldsymbol{\eta}}, b, \theta_{\epsilon}\}\nonumber\\
&=\{\mu_{\eta_{0}}, \mu_{\eta_{1}}, \psi_{00}, \psi_{01}, \psi_{11}, b, \theta_{\epsilon}\}\nonumber
\end{align}} \\
\item{Jenss-Bayley function: \begin{align}
\boldsymbol{\Theta}_{\text{JB}}&=\{\boldsymbol{\mu}_{\boldsymbol{\eta}}, \boldsymbol{\Psi}_{\boldsymbol{\eta}}, c, \theta_{\epsilon}\}\nonumber\\
&=\{\mu_{\eta_{0}}, \mu_{\eta_{1}}, \mu_{\eta_{2}}, \psi_{00}, \psi_{01}, \psi_{02}, \psi_{11}, \psi_{12}, \psi_{22}, c, \theta_{\epsilon}\}\nonumber
\end{align}}
\end{itemize}

We use the full information maximum likelihood (FIML) technique to estimate each proposed LCSM to account for the heterogeneity of individual contributions to the likelihood. The log-likelihood function of each individual and that of the overall sample are
\begin{equation}\nonumber
\log lik_{i}(\boldsymbol{\Theta}_{\text{LBGM/QUAD/EXP/JB}}|\boldsymbol{y}_{i})=C-\frac{1}{2}\ln|\boldsymbol{\Sigma}_{i}|-\frac{1}{2}\big(\boldsymbol{y}_{i}-\boldsymbol{\mu}_{i})^{T}\boldsymbol{\Sigma}_{i}^{-1}(\boldsymbol{y}_{i}-\boldsymbol{\mu}_{i}\big),
\end{equation}
and
\begin{equation}\nonumber
\log lik(\boldsymbol{\Theta}_{\text{LBGM/QUAD/EXP/JB}})=\sum_{i=1}^{n}\log lik_{i}(\boldsymbol{\Theta}_{\text{LBGM/QUAD/EXP/JB}}|\boldsymbol{y}_{i}),
\end{equation}
respectively, in which $C$ is a constant, $n$ is the number of individuals, $\boldsymbol{\mu}_{i}$ and $\boldsymbol{\Sigma}_{i}$ are the mean vector and the variance-covariance matrix of the longitudinal outcome $\boldsymbol{y}_{i}$. We use the R package \textit{OpenMx} with the optimizer CSOLNP \citep{OpenMx2016package, Pritikin2015OpenMx, Hunter2018OpenMx, User2020OpenMx} to build the proposed models. We provide \textit{OpenMx} code in the online appendix  (\url{https://github.com/Veronica0206/LCSM_projects}) to demonstrate how to employ the proposed novel specification. The proposed LCSMs with the novel specification can also be fit using other SEM software such as \textit{Mplus 8}. We also provide the corresponding code on the GitHub website for researchers who are interested in using it.

In addition to the growth factors, the rate-of-change (i.e., $dy_{ij}$ in the LBGM or $dy_{ij\_\text{mid}}$ in the parametric nonlinear LCSM) and true score (i.e., $y^{\ast}_{ij}$) at each time point are also latent variables in the LCSM framework, as shown in Figure \ref{fig:path}, although the means and variances of them are not free parameters. By using the delta method \citep[Chapter~1]{Lehmann1998Delta}, we are able to derive the mean and variance of $dy_{ij}$ ($dy_{ij\_\text{mid}}$) from Equation \ref{eq:LBGM3} and Equations \ref{eq:quad}-\ref{eq:JB}. The detailed derivation of the mean and variance of the rate-of-change is provided in Appendix \ref{supp:1}. 

Moreover, as stated earlier, other research interests in analyzing longitudinal data include the estimation of the change that occurs during a time interval and the amount of change from baseline. It is straightforward to estimate these values with the proposed model specification since each element of the product of the grey shaded part of each $\boldsymbol{\Lambda}_{i}$ in Equations \ref{eq:LBGM_load}, \ref{eq:quad_load}, \ref{eq:exp_load}, and \ref{eq:JB_load} and the corresponding growth factor(s) is the amount of change-from-baseline at each post-baseline time point. If only the mean values and variances of these parameters are of interest, we can derive them along with the current model specification in SEM software. For example, in \textit{OpenMx}, we can specify the expression for a derived parameter in the function \textit{mxAlgebra()}, and then evaluate the corresponding point estimate and standard error of a derived parameter by using function \textit{mxEval()} and \textit{mxSE()}, respectively. We provide a detailed derivation of the means and variances for the parameters of interval-specific change and change-from-baseline in Appendix \ref{supp:1}. In practice, it may also be of interest to calculate the interval-specific change or change-from-baseline at the individual level. We then need to modify the model specification by adding interval-specific change or the amount of change-from-baseline as latent variables explicitly. Specifically, we need to modify Equation \ref{eq:LBGM1} as
\begin{align}
&y^{\ast}_{ij}=\begin{cases}
\eta_{0i}, & \text{if $j=1$}\\
y^{\ast}_{i(j-1)}+\delta y_{ij}, & \text{if $j=2, \dots, J$}
\end{cases},\label{eq:LCSM_d1}\\
&\delta y_{ij}=dy_{ij}\times(t_{ij}-t_{i(j-1)})\qquad (j=2, \dots, J), \label{eq:LCSM_d2}
\end{align}
where $\delta y_{ij}$ indicates the change occurs between $t=t_{i(j-1)}$ and $t=t_{ij}$, to allow for the estimation of interval-specific changes for the nonparametric LCSM. We are able to estimate interval-specific changes for LCSMs with parametric functional forms by replacing Equation \ref{eq:LCSM_d2} with $\delta y_{ij}\approx dy_{ij\_\text{mid}}\times(t_{ij}-t_{i(j-1)})\quad (j=2, \dots, J)$. Similarly, we need to update Equation \ref{eq:LBGM1} as
\begin{align}
&y^{\ast}_{ij}=\begin{cases}
\eta_{0i}, & \text{if $j=1$}\\
y^{\ast}_{i1}+\Delta y_{ij}, & \text{if $j=2, \dots, J$}
\end{cases},\label{eq:LCSM_D1}\\
&\Delta y_{ij}=\Delta y_{i(j-1)}+dy_{ij}\times(t_{ij}-t_{i(j-1)})\qquad (j=2, \dots, J), \label{eq:LCSM_D2}
\end{align}
where $\Delta y_{ij}$ is the amount of change from baseline at $t=t_{ij}$, to estimate the amount of change-from-baseline for the nonparametric LCSM. Replacing Equation \ref{eq:LCSM_D2} with $\Delta y_{ij}\approx\Delta y_{i(j-1)}+dy_{ij\_\text{mid}}\times(t_{ij}-t_{i(j-1)})\quad (j=2, \dots, J)$ allows us to derive the amount of change from baseline for a parametric LCSM. With \textit{OpenMx} function \textit{mxFactorScores()} \citep{OpenMx2016package, Pritikin2015OpenMx, Hunter2018OpenMx, User2020OpenMx, Estabrook2013score}, we are able to obtain individual values of each of growth factors, rate-of-change, and these additional latent variables. We provide the corresponding code for these possible applications on the GitHub website.

\section*{Model Evaluation}\label{M:Evaluate}
We performed a Monte Carlo simulation study to evaluate the proposed model specification with two objectives. The first objective is to examine the four models with the novel specification introduced in the Method section through performance metrics, including the relative bias, empirical SE, relative RMSE, and empirical CP for a nominal $95\%$ confidence interval of each parameter. In Table \ref{tbl:metric}, we provide the definitions and estimates of these four performance measures. The second objective is to evaluate how the approximated value of the AUC (i.e., the latent change score) in each time interval affects the performance metrics of each nonlinear parametric LCSM. To this end, we generate LGCM-implied data structures for each parametric model, build up the corresponding LGCM and LCSM, and compare the four performance metrics.

\tablehere{2}

In the simulation design, the number of repetitions $S=1,000$ is determined by an empirical method introduced in \citet{Morris2019simulation}. We performed a pilot study and found that the standard errors of all coefficients except the parameters related to the initial status and vertical distance\footnote{In the negative exponential function, the vertical distance is the distance between the initial status and asymptotic level while in the Jenss-Bayley function, it is the distance between the initial status and intercept of the linear asymptote.} were below $0.15$. Therefore, at least $900$ repetitions are needed to keep the Monte Carlo standard error of the bias\footnote{The most important performance metric in a simulation study is the bias, and equation for the Monte Carlo standard error is $\text{Monte Carlo SE(Bias)}=\sqrt{Var(\hat{\theta})/S}$ \citep{Morris2019simulation}.} less than $0.005$. For this reason, we determined to perform the simulation study with $1,000$ replications for more conservative consideration.

\subsection*{Design of Simulation Study}\label{Simu:design}
We provide all the conditions that we considered for each model in the simulation design in Table \ref{tbl:simu_design}. An important factor in models used to investigate longitudinal processes is the number of repeated measures. One hypothesis is that the model performs better as repeated records increase. We were interested in examining this hypothesis through the simulation study. Therefore, we selected two levels of repeated measures for all four models: six and ten. One goal was to assess whether the longitudinal records are equally-placed or not affect the model performance, assuming that we had the same study duration with ten repeated measurements. In addition, we wanted to see how these four models perform under the more challenging condition with shorter study duration and six repeated records. Moreover, we allowed for a `medium' time window $(-0.25, +0.25)$ around each wave, following \citet{Coulombe2015ignoring}. In addition to the time structures, we also considered some same conditions across the four models. For example, we fixed the distribution of the initial status ($\eta_{0i}\sim N(50, 5^2)$) since it only affects the position of a trajectory. In addition, we set the growth factors in each model to be positively correlated to a moderate level ($\rho=0.3$) and considered two levels of sample size ($n=200$ or $500$) and two levels of residual variance ($\theta_{\epsilon}=1$ or $2$) for all four models. 

\tablehere{3}

For the nonparametric LCSM (i.e., LBGM), we examined how the trajectory shape, quantified by the shape factor and relative rate-of-change, affects the model. As shown in Table \ref{tbl:simu_design}, we fixed the distribution of the shape factor and examined the trajectory with a decreasing or increasing rate-of-change in the simulation study. On the other hand, for the exponential LCSM and Jenss-Bayley LCSM, we only considered the nonlinear trajectory with a declining rate-of-change (i.e., the trajectory with an asymptotic level) because identifying the asymptote is one goal of using these parametric LCSMs. We fixed the vertical distance for the negative exponential LCSM to have a constant asymptote level but considered two levels of logarithmic ratio of the growth rate ($b=0.4$ or $0.8$). We fixed the vertical distance and the ratio of the growth acceleration for the Jenss-Bayley LCSM but examined how a different rate-of-change in the later developmental stage, quantified by the slope of the linear asymptote, affects model performance. Specifically, we considered two distributions of the slope of the linear asymptote, $N(2.5, 1.0^{2})$ and $N(1.0, 0.4^{2})$, for a large and small rate-of-change in the later stage. The change of the quadratic function is not monotonic, so we adjusted the linear and quadratic slopes to have a monotonic change of the study with six and ten repeated measurements. For each condition of each model listed in Table \ref{tbl:simu_design}, we carried out the simulation study as the steps described in Appendix \ref{supp:2}.   

\section*{Results}\label{Result}
We first evaluated the convergence\footnote{Convergence in the current project is defined as achieving the \textit{OpenMx} status code $0$ (which suggests that the optimization is successful) until up to $10$ trials with different sets of starting values.} rate of each proposed model and the corresponding LGCM (if applicable). The proposed models and their available LGCM counterparts converged well as they reported a $100\%$ convergence rate for all conditions listed in Table \ref{tbl:simu_design}.

Based on our simulation study, the estimates of all four models with the novel specification were unbiased and accurate, with the target $95\%$ coverage probability in general. Some factors, such as the number of repeated measurements (the length of study duration) and the occasions of these measurements, might affect model performance. Specifically, more measurements, especially more measurements in an earlier stage, could improve the performance of these models. In the simulation study, we found that the negative exponential LGCM and Jenss-Bayley LGCM outperformed the corresponding LCSM, which is within our expectation since we fit the LGCM and LCSM to the corresponding LGCM-implied data structure. Even so, the overall performance of the LCSMs with the novel specification was still satisfactory. The detailed simulation results are provided in Appendix \ref{supp:3}.

\section*{Application}\label{Application}
We now use empirical data to demonstrate how to apply the LCSMs with the novel specification and the corresponding LGCMs (if applicable) to answer research questions. This part of the application has two goals. The first goal is to provide a set of feasible recommendations on how to employ the proposed LCSMs and use the free and derived parameters to answer specific research questions. The second goal is to show how different frameworks with the same trajectory function affect the estimation in this real-world practice; for this reason, we built up the three LGCMs as a sensitivity analysis. In this application, we randomly selected $400$ students from The Early Childhood Longitudinal Study, Kindergarten Class of 2010-2011 (ECLS-K: 2011) with non-missing records of repeated reading assessments and age at each study wave\footnote{There are $n= 18174$ participants in ECLS-K: 2011. After removing rows with missing values (i.e., records with any of $NaN/-9/-8/-7/-1$), we have $n=3418$ students.}. 

ECLS-K: 2011 is a nationwide longitudinal study starting from the 2010-2011 school year and collects records from US children enrolled in approximately $900$ kindergarten programs. In ECLS-K: 2011, the reading ability of students was evaluated in nine waves: each semester in kindergarten, first, and second grade, followed by once a school year (only spring semester) in third, fourth, and fifth grade. In the fall semester of $2011$ and $2012$, only about $30\%$ of students were assessed \citep{Le2011ECLS}. In this analysis, we used the child's age (in years) for each wave so that each student had different measurement times. Table \ref{tbl:raw} shows the mean and standard deviation of the observed item response theory (IRT) scores and the amounts of change-from-baseline of reading ability at each study wave. 

\subsection*{Main Analysis}
We fit the nonparametric and three parametric LCSMs with the novel specification to analyze the development of the reading ability of students. All four models converged within half a minute. We provide the estimated likelihood, AIC, BIC, variance of residual, and the number of parameters of each LCSM in Table \ref{tbl:info}. The table shows that the nonparametric LCSM (i.e., LBGM) outperformed three parametric LCSMs from the statistical perspective since it has the largest estimated likelihood and the smallest values of information criteria, including the AIC and BIC. This is not surprising: the model without a pre-specified functional form better captured the data structure, which, in turn, generated a larger likelihood.

\tablehere{5}

Table \ref{tbl:est_LBGM} presents the estimates of parameters of the LBGM. As introduced earlier, we scaled $\eta_{1}$ as the growth rate in the first time interval of the developmental process of reading ability. That is, the parameters related to the initial status and the growth rate in the first interval as well as the values of relative rate-of-change were directly estimated from the proposed model, while the mean and variance of the absolute rate-of-change during each time interval were derived using the function \textit{mxAlgebra()} with \textit{mxEval()} and \textit{mxSE()}. Furthermore, the estimated variability of the initial status and rate-of-change was significant, suggesting the students had individual intercepts and slopes and, thus, individual growth trajectories. In addition, there was a gradual slowdown in the development of reading skills since the growth rate declined over time, as indicated by the shrinking $\gamma_{j}$'s. For example, $\gamma_{5}=0.660$ suggests that the mean growth in the $5^{th}$ interval was only $66\%$ of the growth rate during the first interval. Specifically, reading ability development slowed down post-Grade $3$ in general. Therefore, it suggests that the parametric functions with an asymptotic level, such as the negative exponential or Jenss-Bayley growth curves, can be employed to identify each student's capacity for reading ability.

\tablehere{6}

The estimates of the parametric LCSMs are summarized in Tables \ref{tbl:est_QUAD}-\ref{tbl:est_JB}. For each parametric functional form, the output of the LCSM includes the estimates of the growth coefficients that are also available in the corresponding LGCM. In addition, we can obtain the estimated mean and variance of the instantaneous slope midway in each time interval as shown in Tables \ref{tbl:est_QUAD}-\ref{tbl:est_JB}. The mean values of rate-of-change were not constant but declined with age. In Table \ref{tbl:est_QUAD}, it is noticed that the rate-of-change of the quadratic LCSM decreased linearly, indicated by the constant negative acceleration. Specifically, $\mu_{\eta_{2}}=-2.489$ suggests that, on average, the change in rate-of-change (i.e., acceleration) of the development of reading ability was $-4.978$ (i.e., $-2.489\times2$) each year. In Table \ref{tbl:est_EXP}, it is observed that the average capacity of reading ability across students was $118.531$, with a decreasing deceleration suggested by the ratio of rate-of-change at $t_{j}$ to that at $t_{j-1}$ was $0.708$ (i.e., $e^{-0.345}$). The estimates from the LCSM with Jenss-Bayley functional form also suggested decreasing deceleration as indicated by $c=-0.318$ in Table \ref{tbl:est_JB}. Specifically, this suggested that the ratio of acceleration at $t_{j}$ to that at $t_{j-1}$ was $0.728$ (i.e., $e^{-0.318}$). In addition, all three parametric LCSMs suggested significant individual differences in the rate-of-change in reading ability development. For the quadratic and Jenss-Bayley LCSM, the variability of the rate-of-change first decreased and then increased, while for the negative exponential LCSM, the variability declined monotonically.

\tablehere{7}

\tablehere{8}

\tablehere{9}

Another important output of the LCSM is the amount of change-from-baseline at each post-baseline time point, which is a commonly used metric to evaluate a change in an observational study or a treatment effect in an intervention. In Figure \ref{fig:plot_CHG}, we plot the model-implied change-from-baseline on the smooth line of the corresponding observed values of the reading IRT scores for each proposed LCSM. It can be seen from the figure that the estimated values of change-from-baseline from all three parametric LCSMs can capture the observed values well.

\subsection*{Sensitivity Analysis}
We built up the corresponding LGCM for each LCSM as a sensitivity analysis. The estimated likelihood, AIC, BIC, and residual variance of each LGCM are also provided in Table \ref{tbl:info}. From the table, we note that the values of AIC and BIC of the LGCM were smaller than the corresponding values of LCSM. In addition, we derived the values of change-from-baseline for the three LGCMs and provided the LGCM-based change-from-baseline in Figure \ref{fig:plot_CHG}. We noticed from the figure that the LGCMs tended to overestimate the amount of change-from-baseline. One possible reason for the poor performance of the LGCM in estimating the amount of change-from-baseline in this application is that all three LGCMs underestimated the intercept means (the estimated mean of the initial status was $38.317$, $32.295$, and $33.586$ from the quadratic, negative exponential, and Jenss-Bayley LGCM, respectively). In the Discussion section, we will further explain the implication of such differences in the information criteria and the estimation of the amount of change-from-baseline. 

\section*{Discussion}\label{Discussion}
This article extends the existing LCSM framework to allow for unstructured measurement occasions. Specifically, we view the growth over time as the AUC of the $r-t$ graph. For the parametric LCSM, we propose to approximate the latent change score (i.e., the AUC) within a time interval as the product of the instantaneous slope midway through the interval and the length of the interval. We examined four LCSMs with the proposed specification through extensive simulation studies. Based on our results, with the novel specification, the nonparametric LCSM is capable of providing unbiased and accurate point estimates with target coverage probabilities. For each parametric LCSM, we generated LGCM-implied data structures and constructed the corresponding LGCM and LCSM. Additionally, we apply the proposed models to analyze the developmental process of reading ability using a subsample of $n=400$ from ECLS-K: $2011$. Based on our examination, the parametric LCSM with the novel specification is an ideal alternative to the corresponding LGCM for two considerations. First, it performs satisfactorily on the corresponding LGCM-specified data structure in general. Second, it is capable of providing more information, such as rate-of-change, interval-specific change, and change-from-baseline, than the corresponding LGCM. Such additional information about change allows one to evaluate a nonlinear longitudinal process holistically.

\subsection*{Practical Considerations}\label{D:practical}
In this section, we provide a set of recommendations for empirical researchers based on the simulation study and real-world data analysis. First, it is not our aim to demonstrate that the LCSM framework is universally preferred, although the proposed novel specification of the LCSM has multiple good features, such as providing information regarding change and allowing for unequally-spaced study waves and individual measurement occasions around each wave. Suppose the research interest only focuses on analyzing the observed longitudinal outcomes. In that case, we recommend using the LGCM because the simulation study has shown that the parametric LGCM slightly outperforms the corresponding LCSM in some challenging conditions. In addition, the model specification of the LGCM is more straightforward, and therefore, the insight interpretation is more explicit. However, suppose the research interest is to examine change, including the rate-of-change, interval-specific change, or the amount of change from baseline, especially the examination of such change at the individual level. In that case, the LCSM framework is a great candidate. As demonstrated in the Application section, the LCSM can also estimate the means and variances of the instantaneous slope over time. This allows one further to examine between-individual differences in within-individual changes in nonlinear trajectories.

Additionally, as shown in the Application section, a parametric LGCM fit the data better from the statistical perspective (i.e., the greater estimated likelihood) but generated poorer estimates of the change-from-baseline than the corresponding LCSM. One possible explanation is that no functional forms we considered in this analysis can perfectly capture the underlying change pattern of the raw trajectories, which is typical in any real-world analysis. Specifically, we utilize a pre-specified functional form to capture the observed longitudinal records when specifying a parametric LGCM. As a result, the estimates of growth coefficients from the LGCM fit the majority of observed values well. For example, the LGCM underestimated the initial status in our case, but the estimated coefficients fit the post-baseline observed values well; therefore, it overestimated the amount of the change-from-baseline. However, when specifying the corresponding LCSM, we employ the first derivative of the function, which is unrelated to the initial status, to constrain the pattern of rate-of-change. To this end, the estimates of growth coefficients from the LCSM fit the observed initial status and the first derivative values well, but not necessarily for the observed measurements.

Moreover, the selection of the time unit affects the estimates of the nonlinear trajectories, especially for the growth coefficient $b$ in the negative exponential function and $c$ in the Jenss-Bayley growth curve, since $b$ and $c$ measure the ratio of the growth rate and the ratio of the growth acceleration at two consecutive time points, respectively. Therefore, these two coefficients vary with the time unit. Suppose we fit a model with the negative exponential function. If we select month as the time unit, the growth coefficient $\exp(-b)$ is interpreted as the ratio of the rate-of-change of a month to its precedent month. Similarly, if we select year as the time unit, it is interpreted as the ratio of the rate-of-change of a year to its precedent year. These two values are often not identical since the month-to-month ratio is not expected to be the same as the year-to-year ratio in developmental theory. For example, the estimated ratio is $0.708$ (i.e., $e^{-0.345}$) when we use age-in-year as the unit, as demonstrated in the Application section. The estimated ratio would be $0.971$ (i.e., $e^{-0.029}$) if we consider age-in-month as the time scale. The estimated $b$ values are $-0.345$ and $-0.029$ for the unit year and month, respectively\footnote{Upon further examination, the relationship between these two estimated ratios is $e^{-0.345}\approx (e^{-0.029})^{12}$.}. In practice, we recommend using a relatively large time unit (for example, age-in-year instead of age-in-month in the Application section) to observe a reasonable effect size and ensure the interpretation of these coefficients is meaningful to empirical studies. 

\subsection*{Methodological Considerations and Future Directions}\label{D:method}
This article introduces a novel specification for the LCSM to allow for individual measurement occasions and demonstrates how to apply this proposed specification to fit the LCSM with nonparametric and parametric functional forms. When fitting the LCSM with the negative exponential and Jenss-Bayley functions, we assume that the growth coefficients $b$ and $c$ are roughly the same across individuals to build a parsimonious model. However, these growth coefficients could also be individually different, as stated earlier. Accordingly, one possible extension is to relax the assumption of the fixed growth rate ratio or growth acceleration ratio and examine their random effects to assess individual-level ratios in the LCSM framework as an application warrants \citep{Liu2022LCSM_JB}. In addition, as stated earlier, \citet{Sterba2014individually} proposed to build up a LBGM in the LGCM framework. The examination of the connection and comparison of it and the proposed nonparametric LCSM could be a future direction.

The novel specification of the latent change score can also be extended to other commonly used LCSMs, such as proportional change models, dual change models, and multivariate LCSMs. Multiple existing studies have demonstrated that SEM software, such as \textit{OpenMx} and \textit{Mplus 8}, allow for the examination of residual covariances for multivariate LGCMs \citep{Liu2021PBLSGM, Liu2021PBLSGMM}. Therefore, \textit{OpenMx} and \textit{Mplus 8}, unlike the \textit{SAS} procedure \textit{NLMIXED} \citep{Grimm2018Individually}, should be capable of estimating residual covariances for multivariate LCSMs. The examination of the performance of multivariate LCSMs using \textit{OpenMx} is out of the scope of the present project, but it can be a future direction. In addition, with the novel specification of the latent change score, we propose a new method to estimate ARCs, more specifically, interval-specific ARCs. For the nonparametric functional form, we propose utilizing the interval-specific slopes, which are constant, as the ARCs. We recommend employing the instantaneous slope midway through an interval for a parametric function to approximate the ARC. Appendix \ref{supp:1} provides detailed derivation for the mean and variance for the ARCs of each functional form. Based on the performance of the simulation study, these ARCs should be generally unbiased.

Moreover, one benefit of the LCSM is that it can estimate the interval-specific change and the amount of the change-from-baseline at each post-baseline time point. In addition to using this metric to evaluate the amount of change in one group, researchers may also be interested in comparing the change of multiple manifested or latent groups. Therefore, it is worth extending the LCSM with the novel specification to the multiple-groups framework or finite mixture modeling framework to examine these between-group differences in the amount of change over time. 

Additionally, as shown in the Application section, the nonparametric LCSM tends to estimate the change-related parameters and capture the data structure well but fails to provide coefficients related to developmental theory (e.g., an asymptotic level to suggest capacity). On the contrary, the parametric LCSMs can estimate the coefficients that allow for making hypotheses, yet they may not capture the data structure well. One possible extension is to develop semi-parametric LCSMs, such as a LCSM with a linear-quadratic piecewise function or linear-negative exponential piecewise function. Last, although we demonstrate the LCSMs with the novel specification with complete longitudinal records, it is possible to extend the current work to address a longitudinal data set with dropouts under the assumption of missing at random thanks to the FIML technique.

\subsection*{Concluding Remarks}\label{D:conclude}
This article views the growth curve as the AUC under the $r-t$ graph and proposes a novel specification for the LCSM with one nonparametric and multiple parametric nonlinear functions. The novel specification allows for unequally-spaced study waves and individual measurement occasions around each wave. Other than the information provided by the LGCM, the LCSM is also capable of estimating the means and variances of the instantaneous slope midway in each time interval and the amount of change-from-baseline at each post-baseline time point. The simulation study and application demonstrate the specification's valuable capabilities of estimating all parameters related to change. Furthermore, as discussed above, the proposed specification can be generalized in practice and further examined in methodology.

\bibliographystyle{apalike}
\bibliography{Extension8}

@article{Biesanz2004Linear,
	author = {Biesanz, J. C. and Deeb-Sossa, N. and Papadakis, A. A. and Bollen, K. A. and Curran, P. J.},
	title = {The role of coding time in estimating and interpreting growth curve models.},
	journal = {Psychological methods},
	volume = {9},
	number = {1},
	pages = {30–52},
	year = {2004},
	url = {https://doi.org/10.1037/1082-989X.9.1.30}
}

@article{Zhang2012LCSM,
	author = {Zhang, Z. and McArdle, J. J. and Nesselroade, J. R.},
	title = {Growth rate models: emphasizing growth rate analysis through growth curve modeling.},
	journal = {Journal of Applied Statistics},
	volume = {39},
	number = {6},
	pages = {1241-1262},
	year = {2012},
	url = {https://doi.org/10.1080/02664763.2011.644528}
}

@article{Grimm2013LCSM,
	author = {Grimm, K. J. and Zhang, Z. and Hamagami, F. and Mazzocco, M.},
	title = {Modeling Nonlinear Change via Latent Change and Latent Acceleration Frameworks: Examining Velocity and Acceleration of Growth Trajectories},
	journal = {Multivariate Behavioral Research},
	volume = {48},
	number = {1},
	pages = {117-143},
	year = {2013},
	url = {https://doi.org/10.1080/00273171.2012.755111}
}

@article{Finkel2003cognitive,
	author = {Finkel, D. and Reynolds, C. and Mcardle, J. and Gatz, M. and L Pedersen, N.},
	year = {2003},
	month = {06},
	pages = {535-550},
	title = {Latent Growth Curve Analyses of Accelerating Decline in Cognitive Abilities in Late Adulthood},
	volume = {39},
	journal = {Developmental psychology},
	url = {https://doi.org/10.1037/0012-1649.39.3.535}
}

@article{Mehta2000people,
	author = {Mehta, P. D. and West, S. G.},
	title = {Putting the individual back into individual growth curves.},
	journal = {Psychological Methods},
	volume = {5},
	number = {1},
	pages = {23-43},
	year  = {2000},
	url = {https://doi.org/10.1037/1082-989x.5.1.23}
}

@book{Grimm2016growth,
    title = {Growth Modeling: Structural Equation and Multilevel Modeling Approaches},
    author = {Grimm, K. J. and Ram, N. and Estabrook, R.},
    year = {2016},
    publisher = {Guilford Press}
}

@article{Kelley2008ARC,
    author = {Kelley, K. and Maxwell, S. E.},
    title = {Delineating the Average Rate of Change in Longitudinal Models.},
    journal = {Journal of Educational and Behavioral Statistics},
    volume = {33},
    number = {3},
    pages = {307–332},
    year = {2008},
    URL = {https://doi.org/10.3102/1076998607306074}
}

@article{Kelley2009ARC,
    author = {Kelley, K.},
    title = {The average rate of change for continuous time models.},
    journal = {Behavior Research Methods},
    volume = {41},
    pages = {268–278},
    year = {2009},
    URL = {https://doi.org/10.3758/BRM.41.2.268}
}

@inproceedings{Le2011ECLS,
	author = {L{\^e}, T. and Norman, G. and Tourangeau, K. and Brick, J. M. and Mulligan, G.},
	title = {Early Childhood Longitudinal Study: Kindergarten Class of 2010-2011 - Sample Design Issues},
	booktitle = {JSM Proceedings 2011},
	pages = {1629-1639},
	publisher = {American Statistical Association},
	address = {Alexandria, VA},
	year = {2011},
	url = {http://www.asasrms.org/Proceedings/y2011/Files/301090_66141.pdf}
}

@article{Blozis2008coding,
	author = {Blozis, S. A. and Cho, Y.},
	title = {Coding and Centering of Time in Latent Curve Models in the Presence of Interindividual Time Heterogeneity},
	journal = {Structural Equation Modeling: A Multidisciplinary Journal},
	volume = {15},
	number = {3},
	pages = {413-433},
	year  = {2008},
	publisher = {Routledge},	
	url = {https://doi.org/10.1080/10705510802154299}
}

@article{Coulombe2015ignoring,
	author = {Coulombe, P. and Selig, J. P. and Delaney, H. D.},
	title = {Ignoring individual differences in times of assessment in growth curve modeling},
	journal = {International Journal of Behavioral Development},
	pages = {76-86},
	volume = {40},
	number = {1},
	year = {2015},
	url = {https://doi.org/10.1177/0165025415577684}
}

@article{Harville1977linearmixed,
	author = {Harville, D. A.},
	title = {Maximum Likelihood Approaches to Variance Component Estimation and to Related Problems},
	journal = {Journal of the American Statistical Association},
	volume = {72},
	number = {358},
	pages = {320-338},
	year  = {1977},
	publisher = {Taylor \& Francis},
	url = {https://doi.org/10.2307/2286796}
}

@article{Lindstrom1990nonlinearmixed,
	author = {Lindstrom, M. J. and Bates, D. M.},
	title = {Nonlinear Mixed Effects Models for Repeated Measures Data},
	journal = {Biometrics},
	volume = {46},
	number = {3},
	pages = {673-687},
	year = {1990},
	publisher = {Wiley, International Biometric Society},
	url = {https://doi.org/10.2307/2532087}
}

@article{Laird1982random,
 	author = {Laird, N. M. and Ware, J. H.},
 	title = {Random-Effects Models for Longitudinal Data},
 	journal = {Biometrics},
	pages = {963-974},
	publisher = {Wiley, International Biometric Society},
	volume = {38},
 	year = {1982},
 	url = {https://doi.org/10.2307/2529876}
}

@book{hedeker2006longitudinal,
	title = {Longitudinal Data Analysis},
	author = {Hedeker, D. and Gibbons, R. D.},
	series = {Wiley Series in Probability and Statistics},
	year = {2006},
	publisher = {Wiley}
}

@book{Pinheiro1994mixed,
	author = {Pinheiro, J. C.},
	title = {Topics in Mixed Effects Models},
	year = {1994},
	publisher = {University of Wisconsin-Madison}
	
}

@article{Vonesh1992mixed,
	author = {Vonesh, E. F. and Carter, R. L.},
	title = {Mixed-Effects Nonlinear Regression for Unbalanced Repeated Measures},
	journal = {Biometrics},
	number = {1},
	pages = {1-17},
	volume = {48},
	year = {1992},
	publisher = {Wiley, International Biometric Society},
	url = {https://doi.org/10.2307/2532734}
}

@article{Bryk1987hierarchical,
	author = {Bryk, A. S. and Raudenbush, S. W. },
	title = {Application of hierarchical linear models to assessing change.},
	journal = {Psychological Bulletin},
	volume = {101},
	number = {1},
	pages = {147-158},
	year = {1987},
	url = {https://doi.org/10.1037/0033-2909.101.1.147}
}

@article{Ram2007LGC,
	author = {Ram, N. and Grimm, K. J.},
	title = {Using simple and complex growth models to articulate developmental change: Matching theory to method},
	journal = {International Journal of Behavioral Development},
	volume = {31},
	number = {4},
	pages = {303-316},
	year = {2007},
	url = {https://doi.org/10.1177/0165025407077751}
}

@article{Duncan2000behavior,
	author = {Duncan, S. C. and Duncan, T. E. and Strycker, L. A.},
	title = {Risk and protective factors influencing adolescent problem behavior: A multivariate latent growth curve analysis},
	journal = {Annals of Behavioral Medicine},
	pages = {103},
	volume = {22},
	number = {2},
	year = {2000},
	url = {https://doi.org/10.1007/BF02895772}
}

@book{Duncan2013latent,
	author = {Duncan, T. E. and Duncan, S. C. and Strycker, L. A.},
	title = {An Introduction to Latent Variable Growth Curve Modeling: Concepts, Issues, and Application (2nd)},
	publisher = {Routledge},
	year = {2013}
}

@book{Bollen2005LGC,
	author = {Bollen, K. A. and Curran, P. J.},
	title = {Latent Curve Models},
	subtitle = {A Structural Equation Perspective},
	publisher = {John Wiley \& Sons, Inc},
	year = {2005}
}

@article{Bauer2003Estimating,
	author = {Bauer, D. J.},
	title = {Estimating Multilevel Linear Models as Structural Equation Models},
	journal = {Journal of Educational and Behavioral Statistics},
	volume = {28},
	number = {2},
	pages = {135-167},
	year = {2003},
	url = {https://doi.org/10.3102/10769986028002135}
}

@article{Curran2003Multilevel,
	author = {Curran, P. J.},
	title = {Have Multilevel Models Been Structural Equation Models All Along?},
	journal = {Multivariate Behavioral Research},
	volume = {38},
	number = {4},
	pages = {529-569},
	year  = {2003},
	publisher = {Routledge},
	url = {https://doi.org/10.1207/s15327906mbr3804_5}
}

@incollection{Grimm2018Individually,
    title = {Individually Varying Time Metrics in Latent Change Score Models},
    booktitle = {Longitudinal Multivariate Psychology},
    author = {Grimm, K. J. and Jacobucci, R.},
    chapter = {3},
    editor = {Ferrer, E and Boker, S.M. and Grimm, K. J.},
    pages = {61-79},
    year = {2018},
    publisher = {Guilford Press}
}

@article{Liu2022LCSM_JB,
    title = {Jenss–Bayley Latent Change Score Model With Individual Ratio of the Growth Acceleration in the Framework of Individual Measurement Occasions.}, 
    author = {Liu, J},
    journal = {Journal of Educational and Behavioral Statistics},
	year = {2022},
    volume = {47},
	number = {5},
    pages = {507–543},
	url = {https://doi.org/10.3102/10769986221099919}
}

@article{Liu2019BLSGM,
    title = {Obtaining interpretable parameters from reparameterizing longitudinal models: transformation matrices between growth factors in two parameter spaces}, 
    author = {Liu, J. and Perera, R. A. and Kang, L. and Sabo, R. T. and Kirkpatrick, R. M.},
    journal = {Journal of Educational and Behavioral Statistics},
	year = {2022},
    volume = {47},
	number = {2},
    pages = {167–201},
	url = {https://doi.org/10.3102/10769986211052009}
}

@article{Grimm2013LCM1,
    author = {Grimm, K. J. and Zhang, Z. and Hamagami, F. and Mazzocco, M.},
    title = {Modeling Nonlinear Change via Latent Change and Latent Acceleration Frameworks: Examining Velocity and Acceleration of Growth Trajectories},
    journal = {Multivariate Behavioral Research},
    volume = {48},
    number = {1},
    pages = {117-143},
    year  = {2013},
    publisher = {Routledge},
    URL = {https://doi.org/10.1080/00273171.2012.755111}
}

@article{Grimm2013LCM2,
    author = {Grimm, K. J. and Castro-Schilo, L. and Davoudzadeh, P.},
    title = {Modeling intraindividual change in nonlinear growth models with latent change scores.},
    journal = {GeroPsych: The Journal of Gerontopsychology and Geriatric Psychiatry},
    volume = {26},
    number = {3},
    pages = {153–162},
    year = {2013},
    URL = {https://doi.org/10.1024/1662-9647/a000093}
}

@article{Tucker1958functional,
	author = {Tucker, L. R.},
	title = {Determination of parameters of a functional relation by factor analysis},
	journal = {Psychometrika},
	year = {1958},
	month = {03},
	day = {01},
	volume = {23},
	number = {1},
	pages = {19-23},
	url = {https://doi.org/10.1007/BF02288975}
}

@article{Rao1958growth,
	author = {Rao, R. C.},
	title = {Some Statistical Methods for Comparison of Growth Curves},
	journal = {Biometrics},
	year = {1958},
	volume = {14},
	number = {1},
	pages = {1-17},
	publisher = {Wiley, International Biometric Society},
	url = {https://doi.org/10.2307/2527726}
}

@article{Meredith1990latent,
	author = {Meredith, W. and Tisak, J.},
	title = {Latent curve analysis},
	journal = {Psychometrika},
	volume = {55},
	number = {1},
	year = {1990},
	pages = {107-122},
	url = {https://doi.org/10.1007/BF02294746}
}

@article{Grimm2011LBGM,
	author = {Grimm, K. J. and Ram, N. and Hamagami, F.},
	title = {Nonlinear growth curves in developmental research.},
	journal = {Child development},
	volume = {82},
	number = {5},
	year = {2011},
	pages = {1357–1371},
	url = {https://doi.org/10.1111/j.1467-8624.2011.01630.x}
}

@book{Bollen2006LGC,
	author = {Bollen, K. A. and Curran, P. J.},
	title = {Latent Curve Models},
	subtitle = {A Structural Equation Perspective},
	publisher = {John Wiley \& Sons, Inc},
	year = {2006}
}

@article{McArdle1986LBGM,
	author = {McArdle, J. J.},
	title = {Latent variable growth within behavior genetic models.},
	journal = {Behavior Genetics},
	volume = {16},
	number = {1},
	year = {1986},
	pages = {163–200},
	url = {https://doi.org/10.1007/BF01065485}
}

@article{Preacher2015repara,
	author = {Preacher, K. J. and Hancock, G. R.},
	title = {Meaningful aspects of change as novel random coefficients: A general method for reparameterizing longitudinal models},
	journal={Psychological Methods},
	pages={84--101},
	volume={20},
	number={1},
	year={2015},
	url = {https://doi.org/10.1037/met0000028}
}

@article{Sterba2014individually,
	author = {Sterba, S. K.},
	title = {Fitting Nonlinear Latent Growth Curve Models With Individually Varying Time Points},
	journal = {Structural Equation Modeling: A Multidisciplinary Journal},
	volume = {21},
	number = {4},
	pages = {630-647},
	year = {2014},
	publisher = {Routledge},
	url = {https://doi.org/10.1080/10705511.2014.919828}
}

@article{Driver2017ct,
	author = {Driver, C. C. and Oud, J. H. L. and Voelkle, M. C.},
	title = {Continuous Time Structural Equation Modeling with R Package ctsem.},
	journal = {Journal of Statistical Software},
	volume = {77},
	number = {5},
	pages = {1–35},
	year = {2017},
	publisher = {Routledge},
	url = {https://doi.org/10.18637/jss.v077.i05}
}

@article{Driver2018ct,
	author = {Driver, C. C. and Voelkle, M. C. },
	title = {Hierarchical Bayesian continuous time dynamic modeling.},
	journal = {Psychological Methods},
	volume = {23},
	number = {4},
	pages = {774–799},
	year = {2018},
	publisher = {Routledge},
	url = {https://doi.org/10.1037/met0000168}
}

@article{Mehta2005people,
	author = {Mehta, P. D. and Neale, M. C. },
	title = {People are variables too: Multilevel structural equations modeling.},
	journal = {Psychological Methods},
	volume = {10},
	number = {3},
	pages = {259-284},
	year = {2005},
	publisher = {Routledge},
	url = {https://doi.org/10.1037/1082-989X.10.3.259}
}

@incollection{McArdle2001LCSM,
    title = {A latent difference score approach to longitudinal dynamic structural analysis},
    booktitle = {Structural equation modeling: Present and future},
    author = {McArdle, J. J.},
    editor = {Cudeck, R. and du Toit, S. and Sorbom, D.},
    pages = {342–380},
    year = {2001},
    publisher = {Lincolnwood, IL: Scientific Software International}
}

@incollection{McArdle2001LCM1,
    title = {A latent difference score approach to longitudinal dynamic structural analysis.},
    author = {McArdle, J. J.},
    booktitle = {Structural equation modeling: Present and future},
    editor = {Cudeck, R. and du Toit, S. H. C. and Sorbom, D.},
    pages = {342-380},
    year = {2001},
    publisher = {Lincolnwood, IL: Scientific Software International}
}

@incollection{McArdle2001LCM2,
    title = {Latent difference score structural models for linear dynamic analyses with incomplete longitudinal data.},
    author = {McArdle, J. J. and Hamagami, F.},
    booktitle = {Decade of behavior. New methods for the analysis of change},
    editor = {Collins, L. M. and Sayer, A. G.},
    pages = {139–175},
    year = {2001},
    publisher = {American Psychological Association},
    URL = {https://doi.org/10.1037/10409-005}
}

@article{McArdle2009LCM,
	author = {McArdle, J. J.},
	title = {Latent variable modeling of differences and changes with longitudinal data.},
	journal = {Annual review of psychology},
	pages = {577–605},
	volume = {60},
	year = {2009},
	URL = {https://doi.org/10.1146/annurev.psych.60.110707.163612}
}

@article{OpenMx2016package,
	title = {Open{M}x 2.0: {E}xtended structural equation and
	statistical modeling},
	author = {Neale, M. C. and Hunter, M. D. and Pritikin, J. N. and Zahery, M. and Brick, T. R. and 
	Kirkpatrick, R. M. and Estabrook, R. and Bates, T. C. and Maes, H. H. and Boker, S. M.},
	journal = {Psychometrika},
	publisher = {Psychometric Society},
	year = {2016},
	volume = {81},
	number = {2},
	pages = {535-549},
	url = {https://doi.org/10.1007/s11336-014-9435-8}
}

@article{Pritikin2015OpenMx,
	title = {Modular open-source software for {I}tem {F}actor {A}nalysis},
	author = {Pritikin, J. N. and Hunter, M. D. and Boker, S. M.},
	journal = {Educational and Psychological Measurement},
	year = {2015},
	volume = {75},
	number = {3},
	pages = {458-474},
	url = {https://doi.org/10.1177/0013164414554615}
}

@article{Hunter2018OpenMx,
	title = {State Space Modeling in an Open Source, Modular, Structural Equation Modeling Environment},
	author = {Hunter, M. D.},
	journal = {Structural Equation Modeling},
	volume = {25},
	number = {2},
	pages = {307-324},
	year = {2018},
	publisher = {Routledge},
	url = {https://doi.org/10.1080/10705511.2017.1369354}
}

@manual{User2020OpenMx,
	title = {OpenMx 2.17.2 User Guide},
	year = {2020},
	author = {Boker, S. M. and Neale, M. C. and Maes, H. H. and Wilde, M. J. and Spiegel, M. and Brick, T. R. and Estabrook, R. and Bates, T. C. and Mehta, P. and {von Oertzen}, T. and Gore, R. J. and Hunter, M. D. and Hackett, D. C. and Karch, J. and Brandmaier, A. M. and Pritikin, J. N. and Zahery, M. and Kirkpatrick, R. M.},
	url = {https://vipbg.vcu.edu/vipbg/OpenMx2/docs/OpenMx/2.17.2/OpenMxUserGuide.pdf}
}

@book{Lehmann1998Delta,
	author = {Lehmann, E. L. and Casella, G.},
	title = {Theory of Point Estimation, 2nd edition},
	year = {1998},
    publisher = {Springer-Verlag New York, Inc}
}

@article{Estabrook2013score,
	author = {Estabrook, R. and Neale, M.},
	title = {A Comparison of Factor Score Estimation Methods in the Presence of Missing Data: Reliability and an Application to Nicotine Dependence.},
	journal = {Multivariate behavioral research},
	volume = {48},
	number = {1},
	pages = {1–27},
	year  = {2013},
	URL = {https://doi.org/10.1080/00273171.2012.730072}
}

@article{Morris2019simulation,
    author = {Morris, T. P. and White, I. R. and Crowther, M. J.},
    title = {Using simulation studies to evaluate statistical methods},
    journal = {Statistics in Medicine},
    volume = {38},
    number = {11},
    pages = {2074-2102},
    year = {2019},
    url = {https://doi.org/10.1002/sim.8086}
}

@book{Venables2002Statistics,
	title = {Modern Applied Statistics with S},
	author = {Venables, W. N. and Ripley, B. D.},
	publisher = {Springer},
	edition = {Fourth},
	address = {New York},
	year = {2002}
}

@article{Raftery1995BIC,
	author = {Raftery, A.},
	title = {Bayesian Model Selection in Social Research.},
	journal={Sociological Methodology},
	volume = {25},
	pages={111-163},
	year={1995},
	url = {https://doi.org/10.2307/271063}
}

@article{Liu2021PBLSGM,
	title={Estimating Knots and Their Association in Parallel Bilinear Spline Growth Curve Models in the Framework of Individual Measurement Occasions.},
	author={Liu, J. and Perera, R. A.},
	journal = {Psychological Methods (Advance online publication)},
	year={2021},
	url={https://doi.org/10.1037/met0000309}
}

@article{Liu2021PBLSGMM,
	title={Extending growth mixture model to assess heterogeneity in joint development with piecewise linear trajectories in the framework of individual measurement occasions.},
	author={Liu, J. and Perera, R. A.},
	journal = {Psychological Methods (Advance online publication)},
	year={2022},
	url={https://doi.org/10.1037/met0000500}
}

\appendix
\renewcommand{\theequation}{A.\arabic{equation}}
\setcounter{equation}{0}
\renewcommand{\thesection}{Appendix \Alph{section}}
\renewcommand{\thesubsection}{A.\arabic{subsection}}

\section{\textbf{Derivation of Rate-of-Change, Interval-specific Change and Change-from-baseline}}\label{supp:1}
This section provides the detailed derivation for the mean and variance of the rate-of-change ($dy_{ij}$ in the LBGM or $dy_{ij\_\text{mid}}$ in each parametric nonlinear LCSM), interval-specific change ($\delta_{ij}$), and change-from-baseline ($\Delta_{ij}$). 

\subsection{\textbf{Derivation of the mean and variance of Rate-of-Change}}
For the LBGM, we have $dy_{ij}=\eta_{1i}\times\gamma_{j-1}$ ($j=2, \dots, J$) from Equation \ref{eq:LBGM3}. Suppose $f: \mathcal{R}\rightarrow \mathcal{R}$ is a function\footnote{In this project, $f$ is a linear function. Under this scenario, the mean and variance can be derived using the theorem for calculating the mean and variance of linear combinations. The results obtained by the theorem and the delta method are identical.}, which takes a point $\eta_{i}\in\mathcal{R}$ as input and produces $f(\eta_{i})\in\mathcal{R}$ as output. By the delta method, the mean and variance of the rate-of-change of the LBGM can be expressed as $\mu_{dy_{ij}}=\mu_{\eta_{1}}\times\gamma_{j-1}\quad (j=2,\dots,J)$ and $\phi_{dy_{ij}}=\psi_{11}\times\gamma_{j-1}^{2}\quad (j=2,\dots,J)$, respectively. Similarly, the mean and variance of the rate-of-change of each parametric LCSM can be expressed as
\begin{itemize}
\item{Quadratic Function: \begin{align}
&\mu_{dy_{ij\_\text{mid}}}=\mu_{\eta_{1}}+2\times\mu_{\eta_{2}}\times t_{ij\_\text{mid}}\quad (j=2,\dots,J), \nonumber\\
&\phi_{dy_{ij\_\text{mid}}}=\psi_{11}+4\times\psi_{22}\times t^{2}_{ij\_\text{mid}}+4\times\psi_{12}\times t_{ij\_\text{mid}}\quad (j=2,\dots,J), \nonumber
\end{align}} \\
\item{Negative Exponential Function: \begin{align}
&\mu_{dy_{ij\_\text{mid}}}=b\times\mu_{\eta_{1}}\times\exp(-b\times t_{ij\_\text{mid}})\quad (j=2,\dots,J), \nonumber\\
&\phi_{dy_{ij\_\text{mid}}}=\psi_{11}\times[b\times\exp(-b\times t_{ij\_\text{mid}})]^{2}\quad (j=2,\dots,J), \nonumber
\end{align}} \\
\item{Jenss-Bayley function: \begin{align}
&\mu_{dy_{ij\_\text{mid}}}=\mu_{\eta_{1}}+c\times\mu_{\eta_{2}}\times\exp(c\times t_{ij\_\text{mid}})\quad (j=2,\dots,J), \nonumber\\
&\phi_{dy_{ij\_\text{mid}}}=\psi_{11}+\psi_{22}\times[c\times\exp(c\times t_{ij\_\text{mid}})]^{2}+2\times\psi_{12}\times c\times\exp(c\times t_{ij\_\text{mid}})\quad (j=2,\dots,J).\nonumber
\end{align}}
\end{itemize}

As we can see from the above equations, the mean value and variance of the rate-of-change of the LBGM are fixed at each study wave, while these parameters from a parametric LCSM are individual-specific in the framework of individual measurement occasions since they are functions of the middle point of each time interval. Therefore, in practice with individual measurement occasions, one may set $t_{ij\_\text{mid}}$ as the average value of the middle of two consecutive measurement times across all individuals to simplify the calculation of the mean and variance of the rate-of-change for the parametric LCSMs.

\subsection{\textbf{Derivation of the mean and variance of interval-specific change}}
It is straightforward to derive the mean value and variance for each interval-specific change from the corresponding value of rate-of-change. Specifically, the mean and variance of each interval-specific change for the nonparametric LCSM can be expressed as $\mu_{\delta y_{ij}}=\mu_{dy_{ij}}\times(t_{ij}-t_{i(j-1)})\quad (j=2,\dots,J)$ and $\phi_{\delta y_{ij}}=\psi_{11}\times\gamma^{2}_{j-1}\times(t_{ij}-t_{i(j-1)})^{2}\quad (j=2,\dots,J)$, respectively. Similarly, for a parametric LCSM, the mean and variance of each interval-specific change can be expressed as $\mu_{\delta y_{ij}}=\mu_{dy_{ij\_\text{mid}}}\times(t_{ij}-t_{i(j-1)})\quad (j=2,\dots,J)$ and $\phi_{\delta y_{ij}}=\phi_{dy_{ij\_\text{mid}}}\times(t_{ij}-t_{i(j-1)})^{2}\quad (j=2,\dots,J)$, respectively. Similar to the rate-of-change, these values of interval-specific change are also individual-specific in the framework of individual measurement occasions.

\subsection{\textbf{Derivation of the mean and variance of change-from-baseline}}
Based on the parameters related to interval-specific change above, we are able to derive the mean value of change-from-baseline at each post-baseline point. In particular, the mean value of change-from-baseline at each post-baseline $t_{ij}$ for a parametric or nonparametric LCSM can be expressed as $\mu_{\Delta y_{ij}}=\sum_{j=2}^{j}\mu_{dy_{ij\_\text{mid}}}\times(t_{ij}-t_{i(j-1)})$ and $\mu_{\Delta y_{ij}}=\sum_{j=2}^{j}\mu_{dy_{ij}}\times(t_{ij}-t_{i(j-1)})$, respectively. The variance of change-from-baseline at each post-baseline point can be expressed as
\begin{itemize}
\item{Nonparametric Function: \begin{align}
&\phi_{\Delta y_{ij}}=\psi_{11}\times\big(\sum_{j=2}^{j}\gamma_{j-1}\times(t_{ij}-t_{i(j-1)})\big)^{2}\quad (j=2,\dots,J), \nonumber
\end{align}} \\
\item{Quadratic Function: \begin{align}
\phi_{\Delta y_{ij}}&=\psi_{11}\times\big(\sum^{j}_{j=2}(t_{ij}-t_{i(j-1)})\big)^{2}+4\times\psi_{22}\times\big(\sum^{j}_{j=2}t_{ij\_\text{mid}}(t_{ij}-t_{i(j-1)})\big)^{2} \nonumber\\
&+4\times\psi_{12}\times\sum^{j}_{j=2}(t_{ij}-t_{i(j-1)})\times\sum^{j}_{j=2}t_{ij\_\text{mid}}(t_{ij}-t_{i(j-1)})\quad (j=2,\dots,J), \nonumber
\end{align}} \\
\item{Negative Exponential Function: \begin{align}
&\phi_{\Delta y_{ij}}=\psi_{11}\times\big(b\times\sum^{j}_{j=2}\exp(-b\times t_{ij\_\text{mid}})\times(t_{j}-t_{j-1})\big)^{2}\quad (j=2,\dots,J), \nonumber
\end{align}} \\
\item{Jenss-Bayley function: \begin{align}
\phi_{\Delta y_{ij}}&=\psi_{11}\times\big(\sum^{j}_{j=2}(t_{ij}-t_{i(j-1)})\big)^{2}+\psi_{22}\times\big(\sum^{j}_{j=2}c\times\exp(c\times t_{ij\_\text{mid}})\times(t_{ij}-t_{i(j-1)})\big)^{2}\nonumber\\
&+2\times\psi_{12}\times\big(\sum^{j}_{j=2}(t_{ij}-t_{i(j-1)})\big)\times\big(\sum^{j}_{j=2}c\times\exp(c\times t_{ij\_\text{mid}})\times(t_{ij}-t_{i(j-1)})\big)\quad (j=2,\dots,J).\nonumber
\end{align}}
\end{itemize}
The mean values and variances of the change-from-baseline are also individual-specific values as the rate-of-change and interval-specific change.

To summarize, the mean values and variances of rate-of-change, interval-specific change, and change-from-baseline values are individual-specific values due to individual measurement occasions. In practice, there are two ways to summarize these values. First, one may plot the mean value and variance for a latent variable of interest (as we did for the mean values of change-from-baseline in Figure \ref{fig:plot_CHG}). Second, it is also possible to obtain approximated values of the mean and variance of a latent variable at each point by fixing individual time points to a specific value (i.e., the mean time point of a study wave across all individuals), as we did for the means and variances of rate-of-change in Tables \ref{tbl:est_QUAD}-\ref{tbl:est_JB}.

\section{Data Generation and Simulation Step}\label{supp:2}
For each condition of each model listed in Table \ref{tbl:simu_design}, we carried out the simulation study according to the following steps:
\begin{enumerate}
\item Generate growth factors for the LBGM and each parametric LGCM using the R package \textit{MASS} \citep{Venables2002Statistics},
\item Generate the time structure with $J$ waves $t_{j}$ as specified in Table \ref{tbl:simu_design} and allow for disturbances around each wave $t_{ij}\sim U(t_{j}-\Delta, t_{j}+\Delta)$ ($\Delta=0.25$) to have individual measurement occasions,
\item Calculate factor loadings of each individual for the LBGM and each parametric LGCM, which are functions of the individual measurement occasions and the additional growth coefficient(s) (if applicable),
\item Calculate the values of the repeated measurements based on the growth factors, factor loadings, and residual variance,
\item Implement each LCSM and the corresponding LGCM (if applicable), estimate the parameters, and construct the corresponding $95\%$ Wald confidence intervals,
\item Repeat the above steps until achieving $1,000$ convergent solutions.
\end{enumerate}

\renewcommand{\thesubsection}{C.\arabic{subsection}}
\setcounter{subsection}{0}

\section{Detailed Simulation Results}\label{supp:3}
\subsection{Performance Metrics}
In this section, we examine the performance metrics of each parameter under all conditions for each LCSM, including relative bias, empirical SE, relative RMSE, and empirical coverage of the nominal $95\%$ confidence interval. For each parameter of each model, we calculated each performance measure across $1,000$ replications under each condition and summarized the values of each performance metric across all conditions into the corresponding median and range. In general, the models with the novel specification are capable of providing unbiased and accurate point estimates with target coverage probabilities. We provide these summaries in Tables \ref{tbl:Metric_LBGM}-\ref{tbl:Metric_JB}.

\tablehere{C1}

\tablehere{C2}

\tablehere{C3}

\tablehere{C4}

Table \ref{tbl:Metric_LBGM} provides the summary of the nonparametric LCSM with the proposed specification. The LBGM was able to generate unbiased point estimates and small empirical SEs. The magnitude of the relative biases of all parameters in the model was less than $0.02$. In addition, except for the parameters related to the intercept, the magnitude of the parameters' empirical SEs was less than $0.19$ (the empirical SE of the intercept mean and variance were below $0.38$ and $2.67$, respectively, while the empirical SE of the covariance between the intercept and shape factor was below $0.43$). In addition, the estimates from the LBGM were accurate: the magnitude of the relative RMSE of all parameters was lower than $0.29$. In addition, the CPs of all parameters under all conditions that we considered in the simulation were around $0.95$, indicating that the $95\%$ confidence interval generated by the LBGM covered the population value at the target level under each condition. 

Table \ref{tbl:Metric_QUAD} provides the summary of the four performance metrics for the quadratic LCSM with the proposed model specification. The performance of the quadratic LCSM was also satisfactory. Specifically, the magnitude of the relative biases and the relative RMSE of all parameters was less than $0.03$ and $0.53$, respectively. Moreover, except for the parameters related to the intercept, the magnitude of the empirical SE was below $0.25$. Additionally, the CPs of all parameters under all conditions were sufficiently close to $0.95$.

Table \ref{tbl:Metric_EXP} lists the summary of the performance metrics of the negative exponential LCSM with the proposed specification. In general, the model performed satisfactorily. In particular, the magnitude of the relative biases of most parameters was below $0.01$, although the relative bias of the mean and variance of the vertical distance was slightly larger, reaching $0.03$ and $0.06$, respectively. In addition, we noticed that the CP of the mean value of the vertical distance was not satisfied. Through further investigation,  we found that the negative exponential LCSM worked better under the conditions with the small ratio of the growth rate (i.e., $b=0.4$) and the long study duration with more records in the early stage (i.e., ten repeated measurements with unequally-spaced waves).

We provide the summary of the performance metrics for the Jenss-Bayley LCSM with the proposed specification in Table \ref{tbl:Metric_JB}. The performance of this model was generally satisfactory. It can be seen from the table that the estimates of the parameters related to the asymptotic slope and vertical distance were not ideal, but they were still acceptable (i.e., the relative bias of these estimates was still less than $10\%$). With further examination, we noticed that the Jenss-Bayley LCSM performed better under the conditions with more repeated measurements, the time structure of unequally-spaced study waves, and the larger sample size.

\subsection{Model Comparison}
This section compares the model performance between each parametric LCSM and the corresponding LGCM based on the four measures and information criteria, including Akaike's Information Criteria (AIC) and Bayesian Information Criteria (BIC). We provide the summary of the performance metrics of quadratic, exponential, and Jenss-Bayley LGCM in Tables \ref{tbl:Metric_QUAD}, \ref{tbl:Metric_EXP}, and \ref{tbl:Metric_JB}, respectively. Under each condition, we noticed that the point estimates of each replication's quadratic LCSM and LGCM stayed consistent up to the fourth decimal place. Therefore, except for one cell, the summary tables of the performance metrics of the two models were the same. Additionally, the estimated likelihood values of the two models were the same in all replications under all conditions up to four decimals, and so were the AIC and BIC. This is what we expect: as shown in Equation \ref{eq:quad} and Figure \ref{fig:rate_quad}, the rate-of-change of the quadratic function has a linear relationship with the time $t$, so the instantaneous slope halfway through a time interval is identical to the ARC in that period. To this end, the estimates of the quadratic LCSM are the same as those from the corresponding LGCM. 

For the negative exponential function and Jenss-Bayley function, the LGCM outperformed the corresponding LCSM, especially in terms of parameter estimation related to the vertical distance (and the slope of linear asymptote of the Jenss-Bayley function). Specifically, although these point estimates from the LGCMs and LCSMs can be considered unbiased (i.e., relative biases are less than $10\%$), the bias from the LGCMs was still relatively smaller. As shown in Tables \ref{tbl:Metric_EXP} and \ref{tbl:Metric_JB}, the negative exponential LCSM and Jenss-Bayley LCSM tended to overestimate\footnote{By `overestimate', we mean that the point estimate is farther from zero than the population value. Similarly, by `underestimate', we mean that the point estimate is closer to zero than the true value.} the vertical distance and slightly underestimate the additional coefficient (i.e., $b$ or $c$). In addition, the Jenss-Bayley LCSM could underestimate the slope of the linear asymptote. This is not surprising. The negative exponential and Jenss-Bayley functions are growth curves with decreasing deceleration (i.e., negative acceleration). Therefore, the instantaneous slope halfway through a period is numerically smaller than the ARC in the time interval. For this reason, both models are likely to overestimate $d_{ij\_{\text{mid}}}$ to satisfy the specified functions. For the negative exponential LCSM, this results in an overestimated vertical distance\footnote{As shown in Equation \ref{eq:exp}, $d_{ij\_{\text{mid}}}$ of the negative exponential growth curve is determined by the vertical distance and the coefficient $b$. Slightly underestimating the coefficient $b$ (such as the relative bias below $0.01$ in our LCSM case) does not affect the estimation of $d_{ij\_{\text{mid}}}$ numerically. Therefore, the overestimation of $d_{ij\_{\text{mid}}}$ leads to an overestimated vertical distance.}.

As shown in Equation \ref{eq:JB}, $d_{ij\_{\text{mid}}}$ of the Jenss-Bayley function consists of the linear asymptote slope and negative exponential term, and the estimation of the two terms is supposed to be complementary. So for the Jenss-Bayley function, an overestimated vertical distance led to an underestimated linear slope. In addition, since these point estimates from the LGCM were less biased, the CPs generated by the LGCM covered the population values better than the corresponding CPs from the LCSM. In addition, among all $24$ conditions of the negative exponential (Jenss-Bayley) function, there were $13$ ($8$) conditions where the difference in the BIC across all replications is less than $6$\footnote{\citet{Raftery1995BIC} has shown that a difference in BIC below $6$ does not suggest strong evidence regarding the model preference.}. For the remaining conditions, at least $63.0\%$ ($87.9\%$) replications reported a BIC difference of less than $6$. It suggests that there is no strong evidence for model preference between the LCSM and the corresponding LGCM under most replications for the simulation study.  

\newpage


\renewcommand\thefigure{\arabic{figure}}
\setcounter{figure}{0}

\begin{figure}
\centering
\begin{subfigure}{.5\textwidth}
  \centering
  \includegraphics[width=1\linewidth]{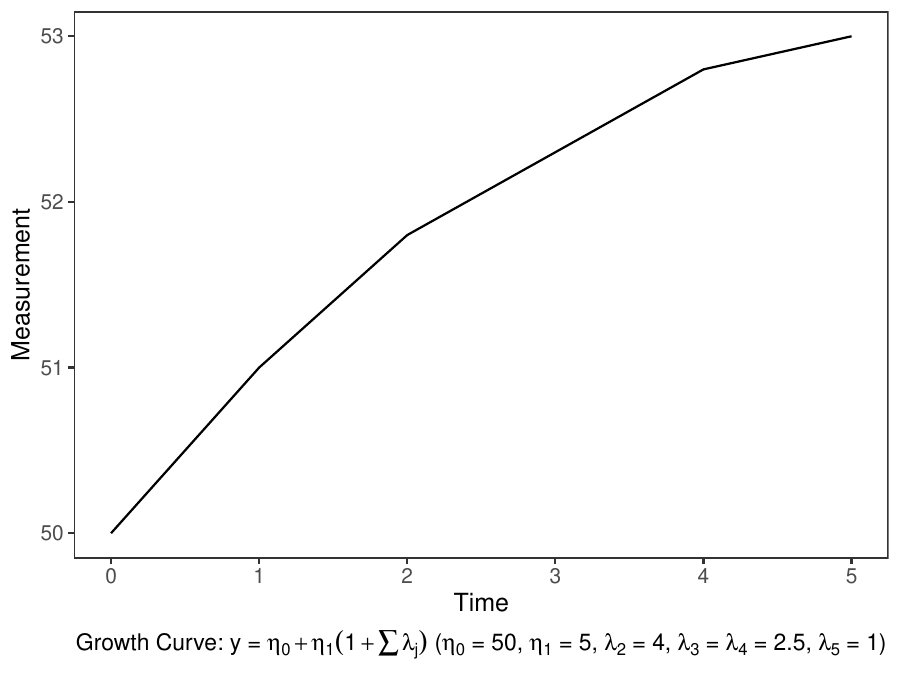}
  \caption{Nonparametric Function}
  \label{fig:growth_nonp}
\end{subfigure}%
\begin{subfigure}{.5\textwidth}
  \centering
  \includegraphics[width=1\linewidth]{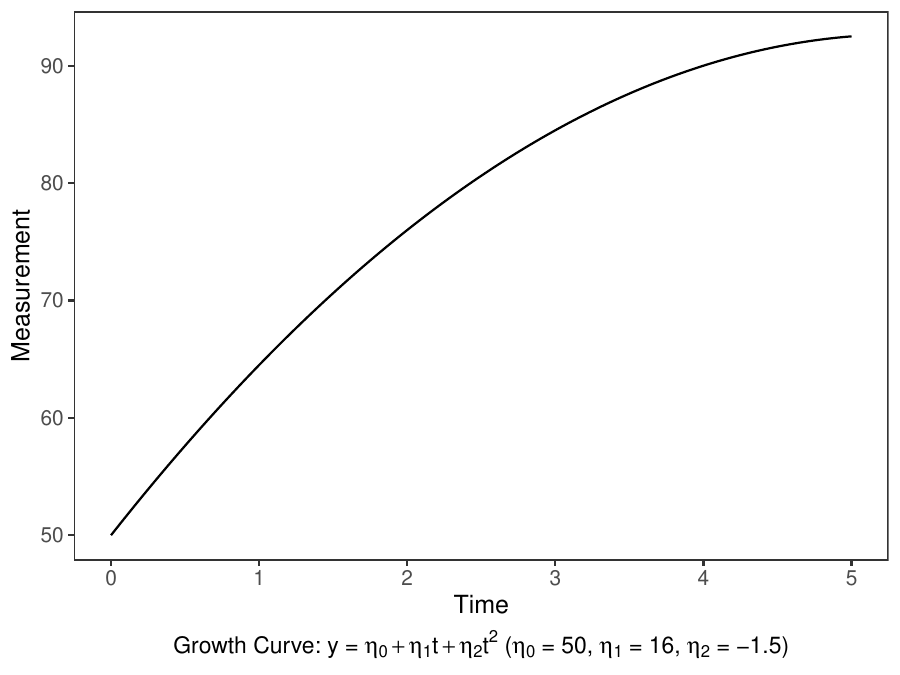}
  \caption{Quadratic Function}
  \label{fig:growth_quad}
\end{subfigure}
\vskip\baselineskip
\begin{subfigure}{.5\textwidth}
  \centering
  \includegraphics[width=1\linewidth]{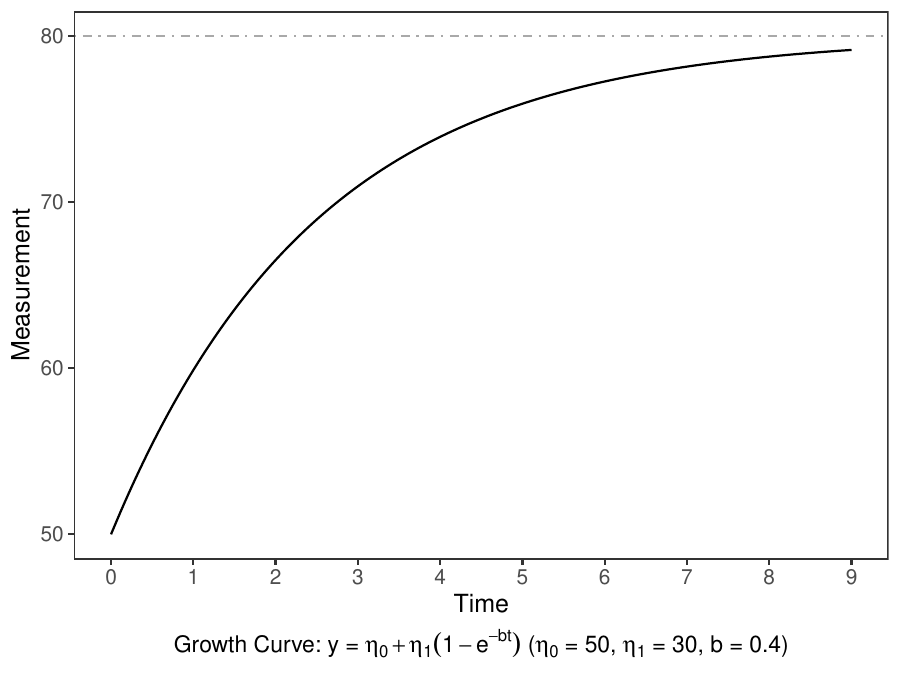}
  \caption{Negative Exponential Function}
  \label{fig:growth_exp}
\end{subfigure}%
\begin{subfigure}{.5\textwidth}
  \centering
  \includegraphics[width=1\linewidth]{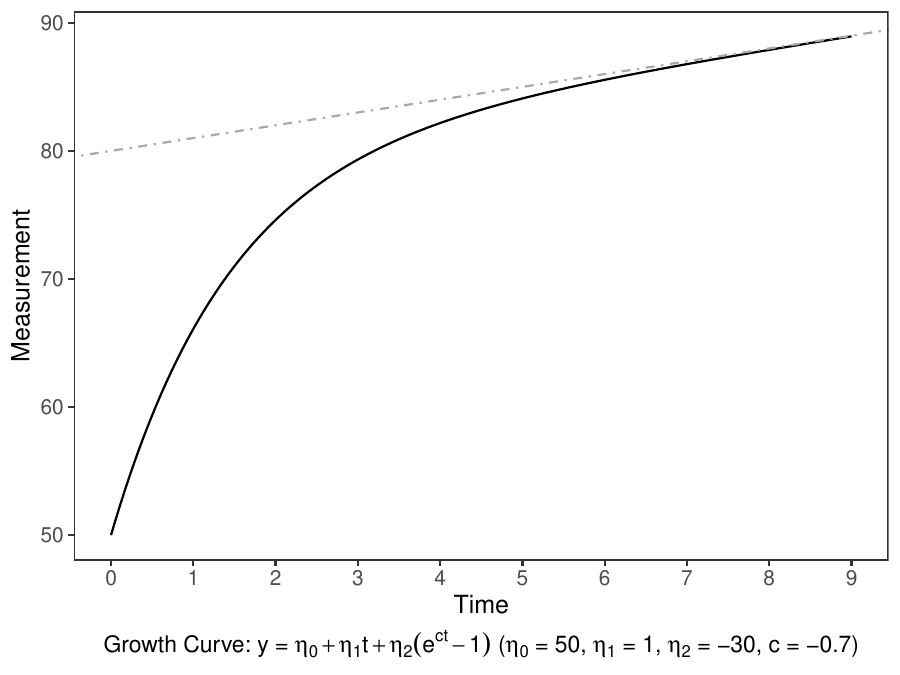}
  \caption{Jenss-Bayley Function}
  \label{fig:growth_JB}
\end{subfigure}
\caption{Growth versus Time Graph of Latent Growth Curve Models}
\label{fig:growth}
\end{figure}

\begin{figure}
\centering
\begin{subfigure}{.5\textwidth}
  \centering
  \includegraphics[width=1\linewidth]{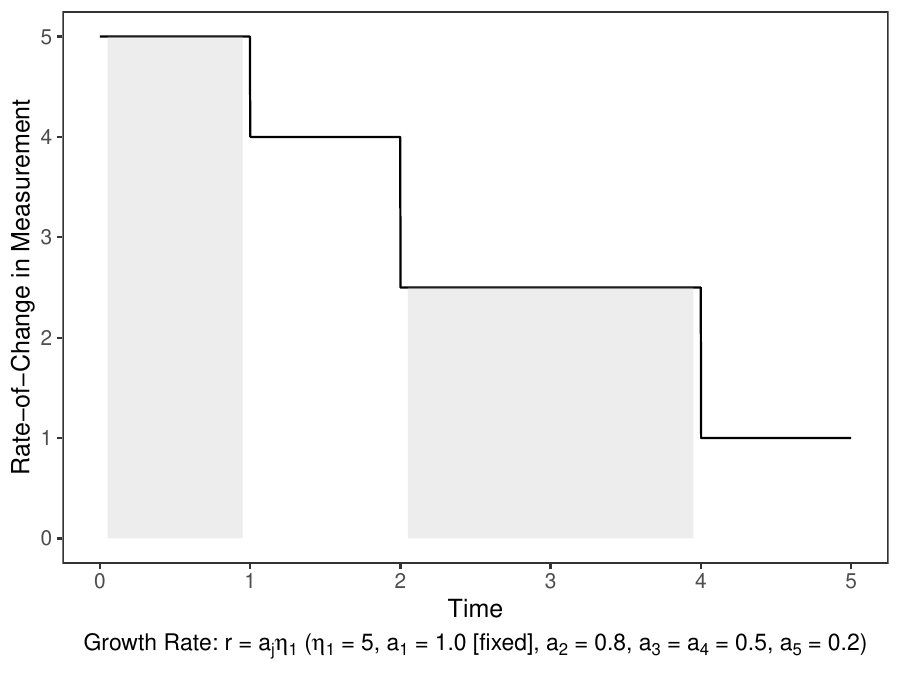}
  \caption{Nonparametric Function}
  \label{fig:rate_nonp}
\end{subfigure}%
\begin{subfigure}{.5\textwidth}
  \centering
  \includegraphics[width=1\linewidth]{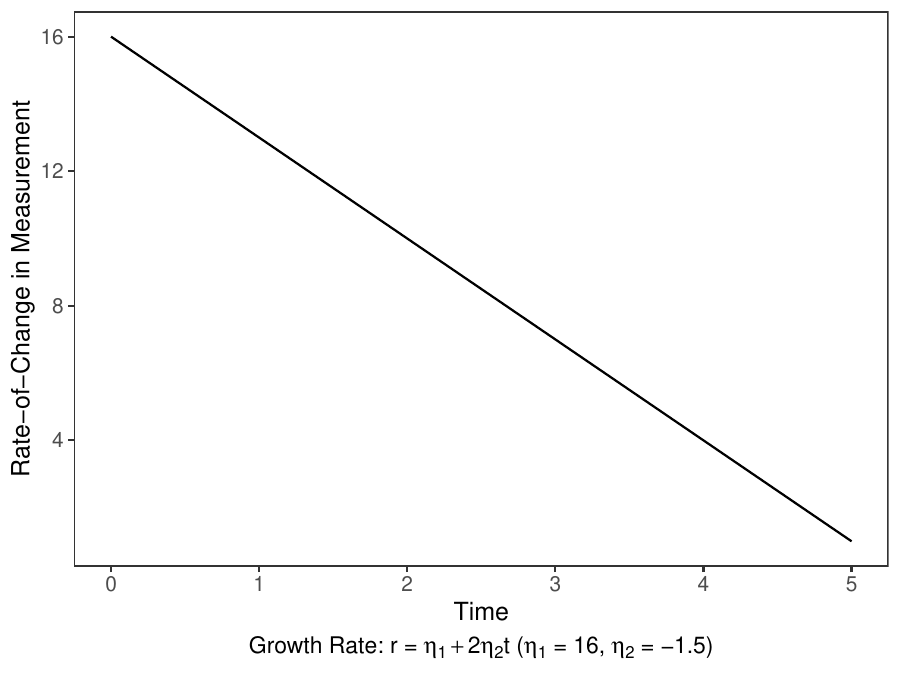}
  \caption{Quadratic Function}
  \label{fig:rate_quad}
\end{subfigure}
\vskip\baselineskip
\begin{subfigure}{.5\textwidth}
  \centering
  \includegraphics[width=1\linewidth]{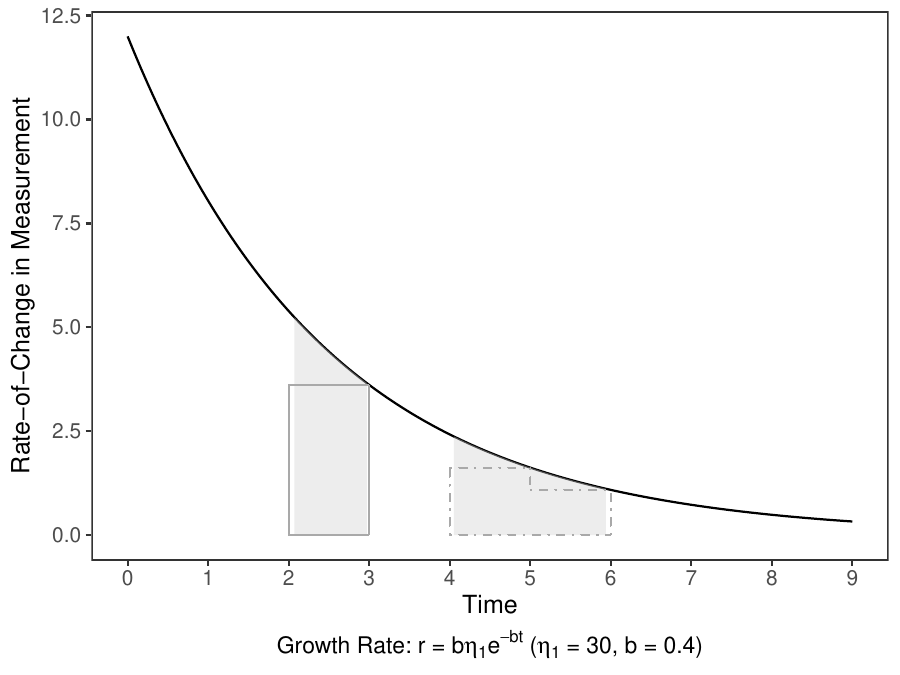}
  \caption{Negative Exponential Function}
  \label{fig:rate_exp}
\end{subfigure}%
\begin{subfigure}{.5\textwidth}
  \centering
  \includegraphics[width=1\linewidth]{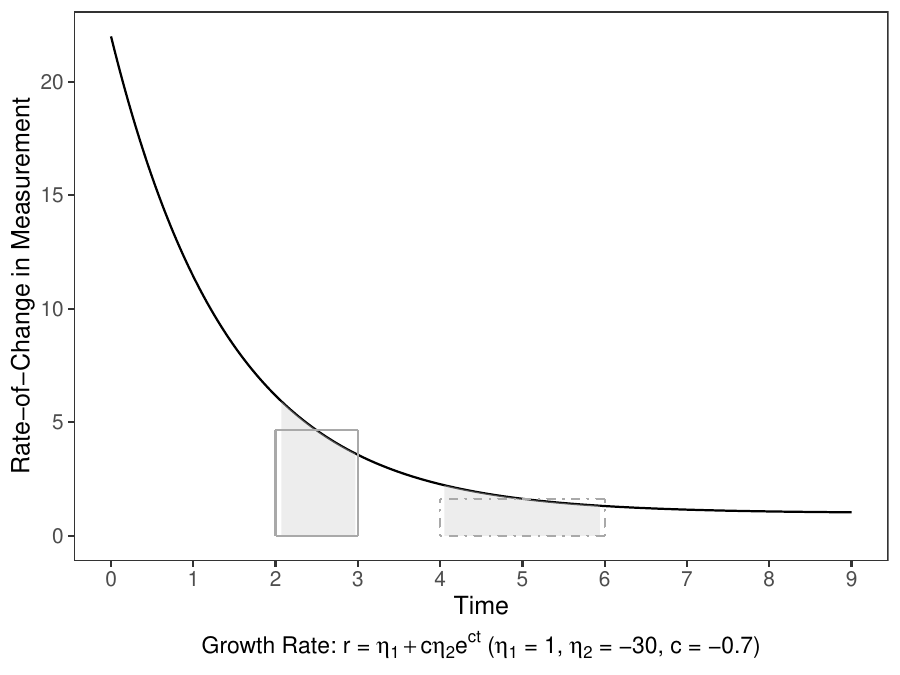}
  \caption{Jenss-Bayley Function}
  \label{fig:rate_JB}
\end{subfigure}
\caption{Rate-of-Change versus Time Graph of Latent Change Score Models}
\label{fig:rate}
\end{figure}

\begin{figure}
\centering
\begin{subfigure}{.5\textwidth}
  \centering
  \includegraphics[width=1\linewidth]{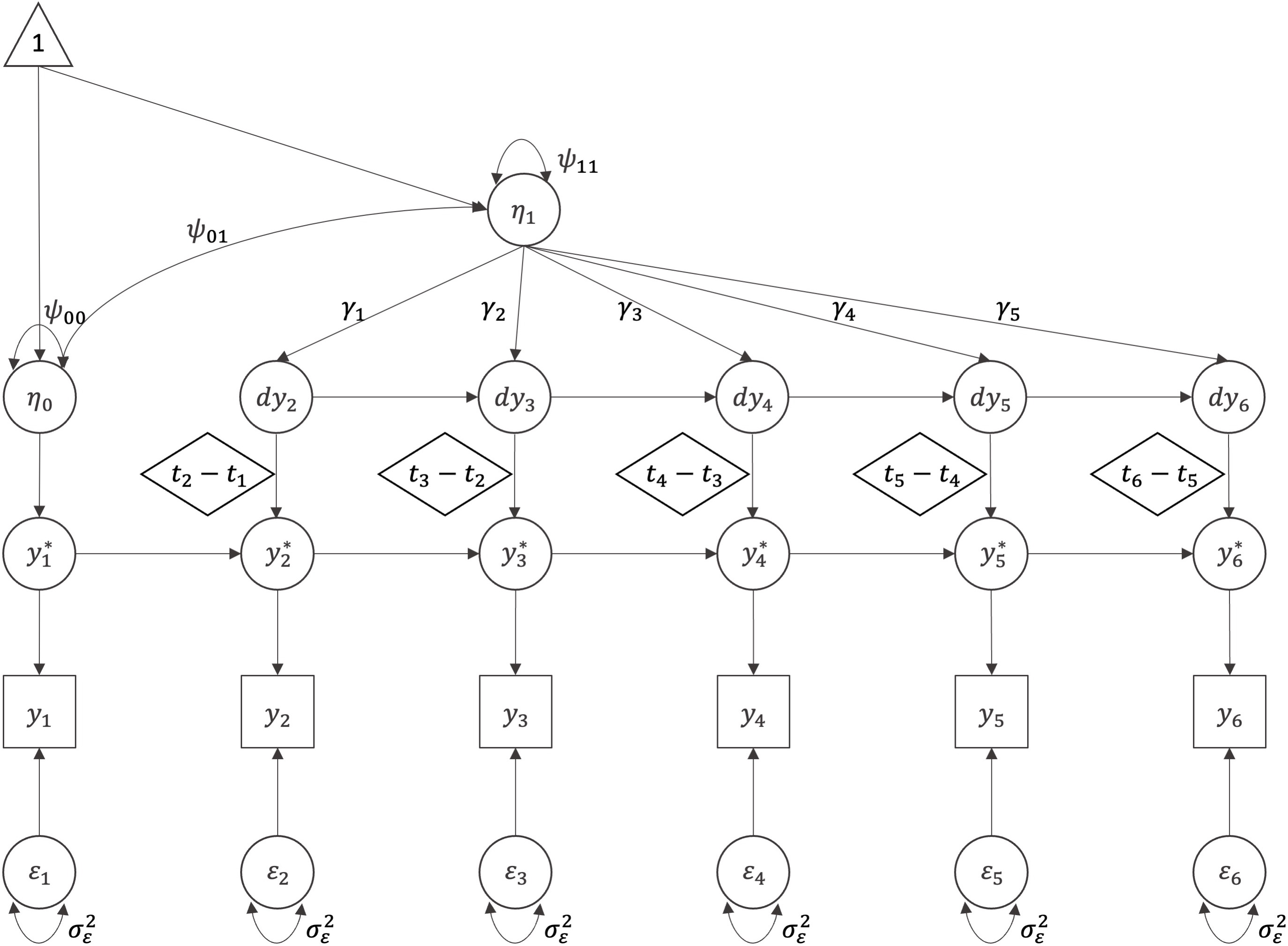}
  \caption{Nonparametric Function}
  \label{fig:path_nonp}
\end{subfigure}%
\begin{subfigure}{.5\textwidth}
  \centering
  \includegraphics[width=1\linewidth]{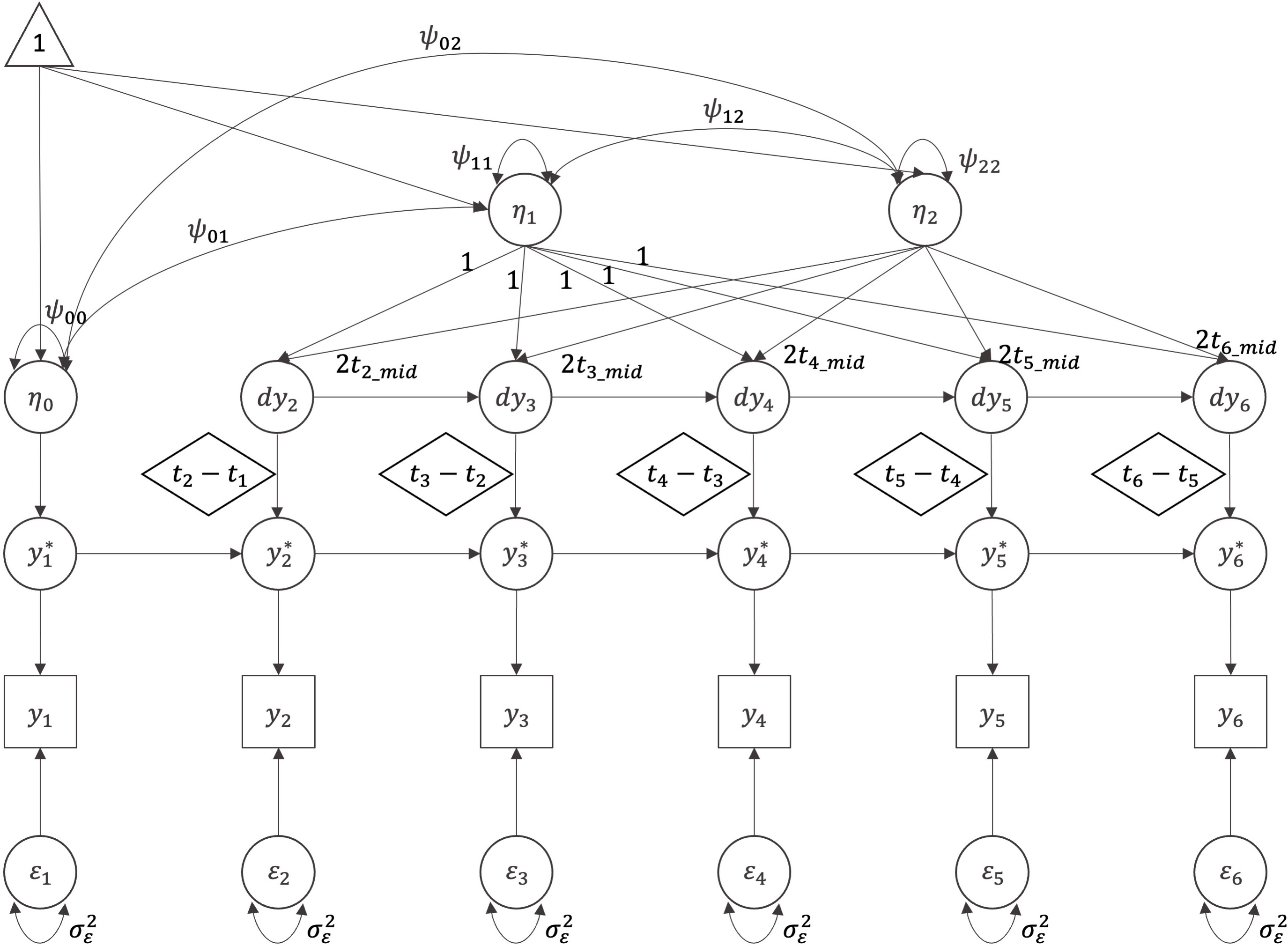}
  \caption{Quadratic Function}
  \label{fig:path_quad}
\end{subfigure}
\vskip\baselineskip
\begin{subfigure}{.5\textwidth}
  \centering
  \includegraphics[width=1\linewidth]{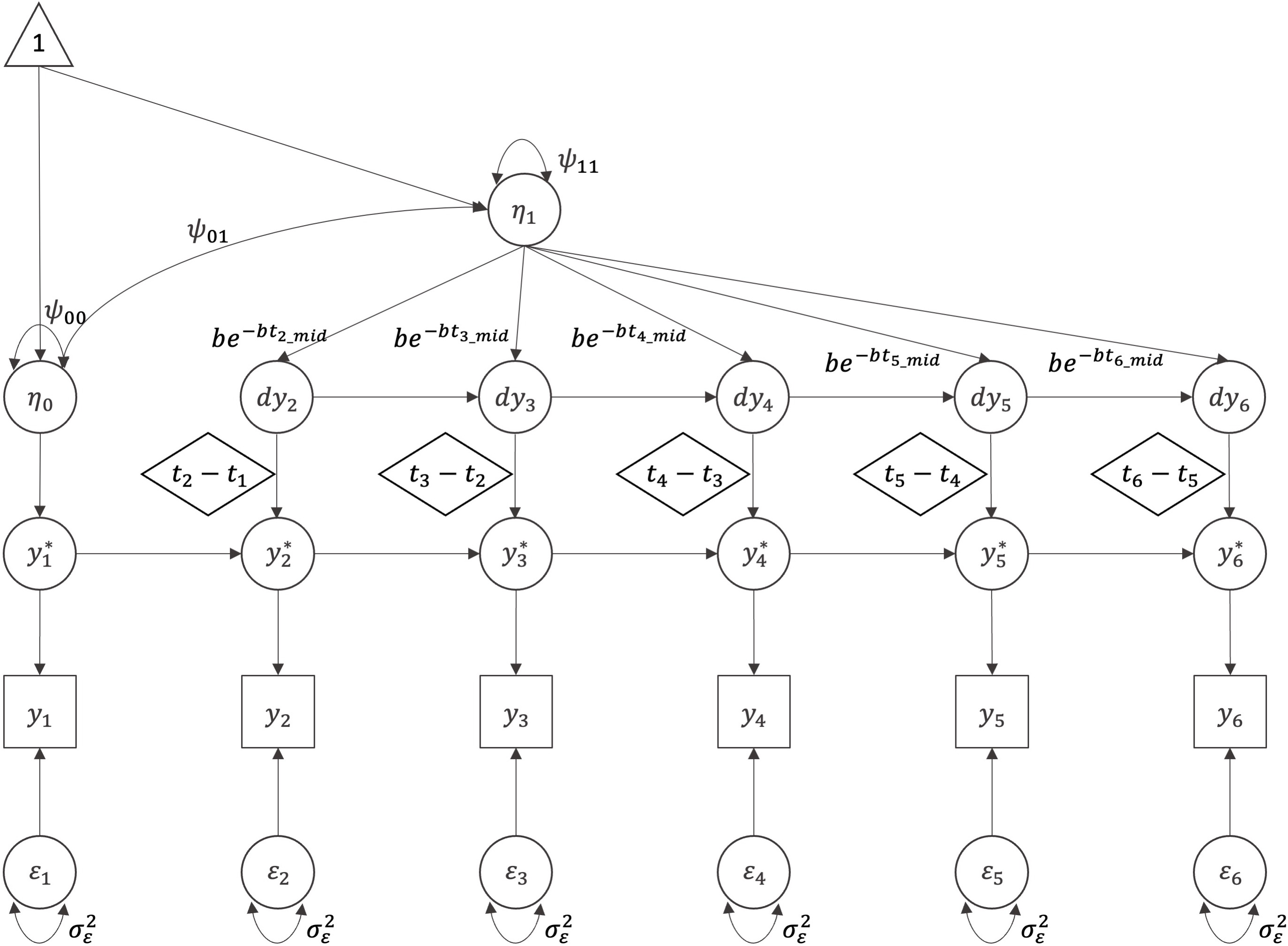}
  \caption{Negative Exponential Function}
  \label{fig:path_exp}
\end{subfigure}%
\begin{subfigure}{.5\textwidth}
  \centering
  \includegraphics[width=1\linewidth]{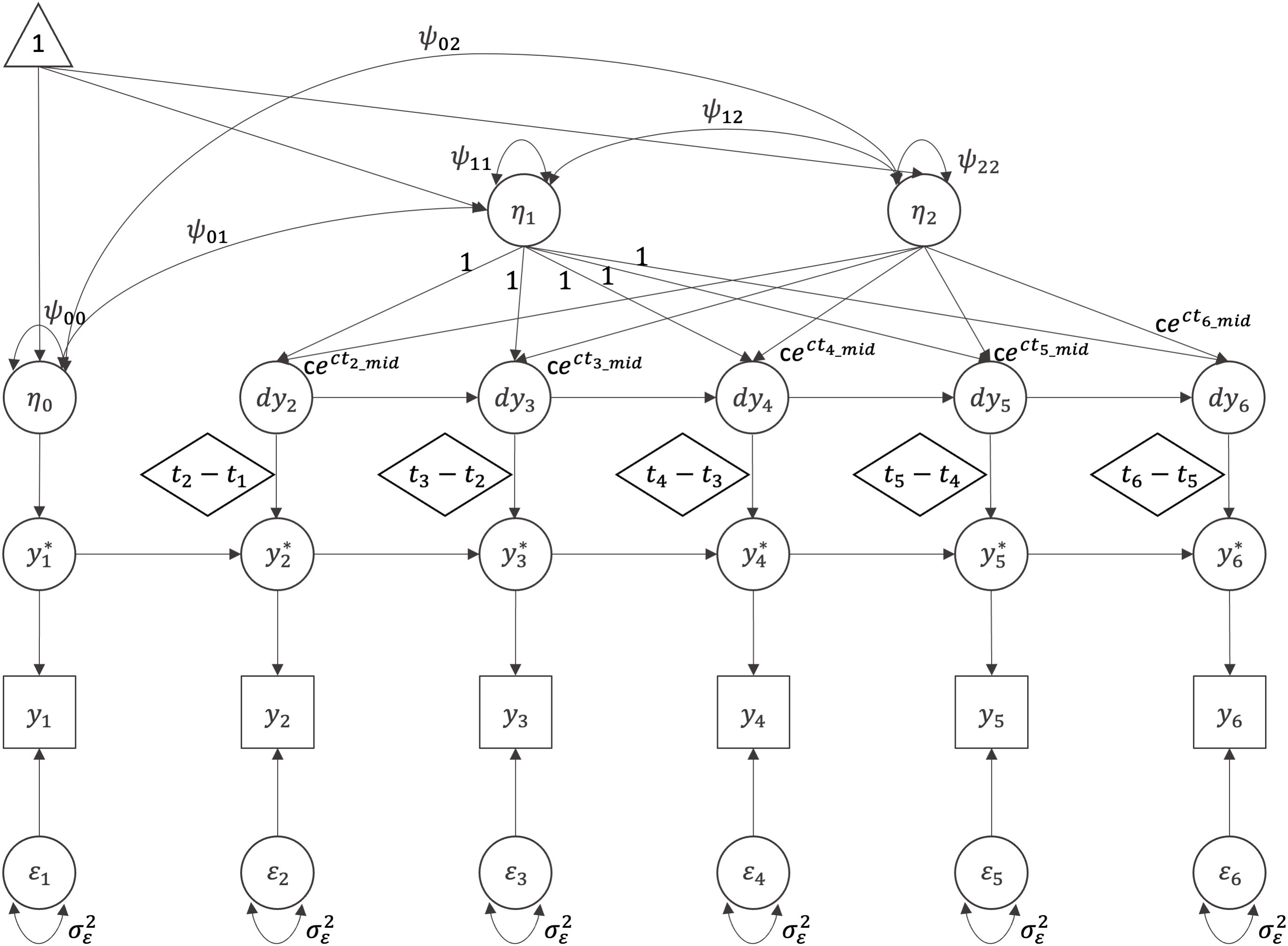}
  \caption{Jenss-Bayley Function}
  \label{fig:path_JB}
\end{subfigure}
\caption{Path Diagram of Latent Change Score Models with the Novel Specification\\
Note: boxes=manifested variables, circles=latent variables, single arrow=regression paths;
doubled arrow=(co)variances; triangle=constant; diamonds=definition variables.\\
In Figure \ref{fig:path_nonp}, we set $\gamma_{1}=1$ for model identification considerations.}
\label{fig:path}
\end{figure}

\begin{figure}
\centering
    \includegraphics[width=1\linewidth]{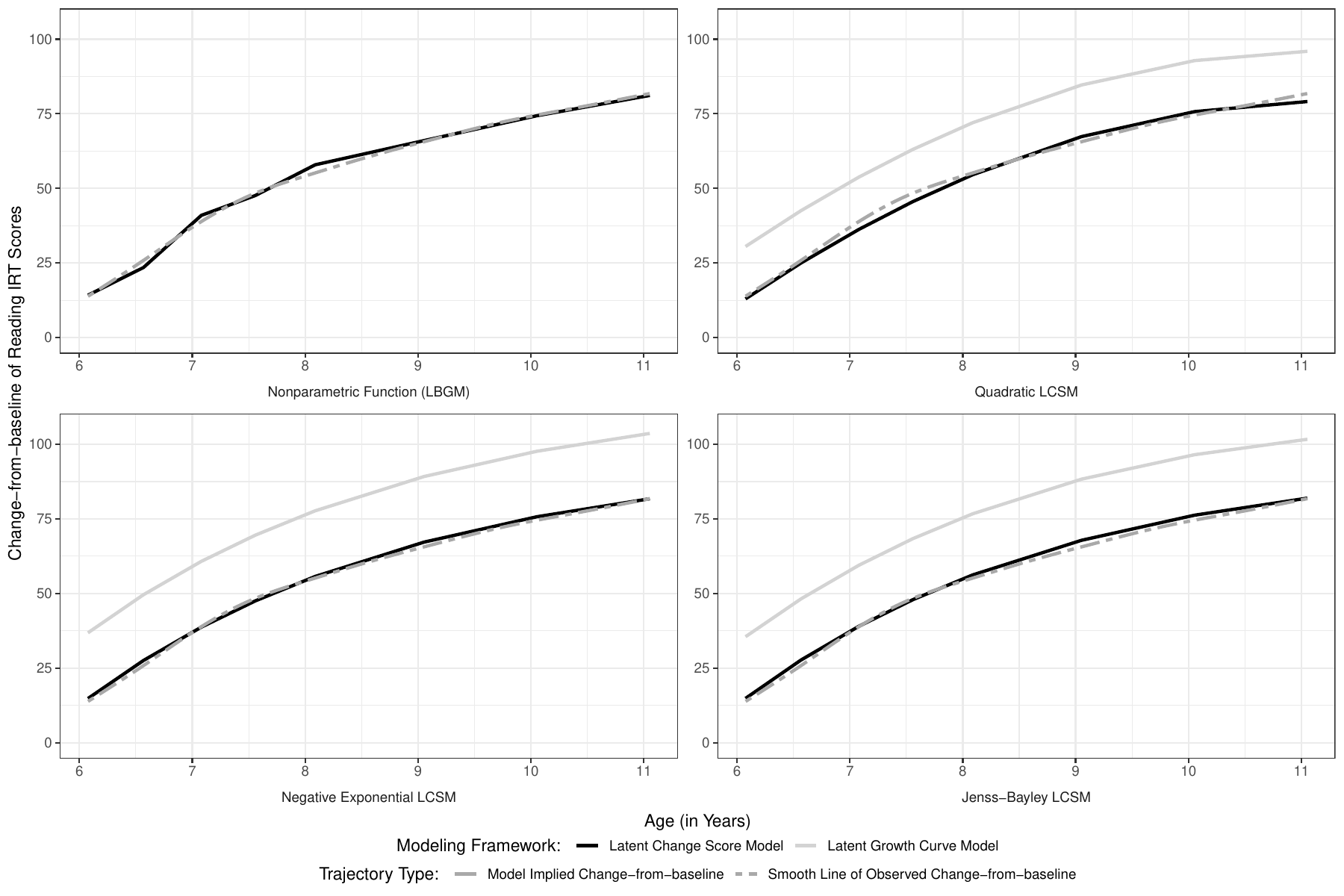}
    \caption{Model Implied Change-from-baseline and Smooth Line of Observed Change-from-baseline}
    \label{fig:plot_CHG}
\end{figure}

\renewcommand\thetable{\arabic{table}}
\setcounter{table}{0}

\begin{table}
\centering
\footnotesize
\rotatebox{90}{%
\begin{threeparttable}
\setlength{\tabcolsep}{5pt}
\renewcommand{\arraystretch}{0.75}
\caption{Summary of Commonly Used nonlinear Functional Forms}
\begin{tabular}{p{7.5cm}p{5.0cm}p{13.0cm}}
\hline
\hline
\multicolumn{3}{c}{\textbf{Quadratic function}}\\
\hline
\textbf{Growth Status} & \textbf{Growth Rate} & \textbf{Coef. related to developmental theory} \\
\hline
\multirow{3}{*}{$y_{ij}=\eta_{0i}+\eta_{1i}\times{t_{j}}+\eta_{2i}\times{t^{2}_{j}}+\epsilon_{ij}$} & \multirow{3}{*}{$d_{ij}=\eta_{1i}+2\times\eta_{2i}\times{t_{j}}$} & $\eta_{0i}$: the initial status \\ & & $\eta_{1i}$: the linear component of change \\ & & $\eta_{2i}$: the quadratic component of change \\
\hline
\hline
\multicolumn{3}{c}{\textbf{Negative exponential function}}\\
\hline
\textbf{Growth Status} & \textbf{Growth Rate} & \textbf{Coef. related to developmental theory} \\
\hline
\multirow{3}{*}{$y_{ij}=\eta_{0i}+\eta_{1i}\times(1-\exp(-b\times {t_{j}}))+\epsilon_{ij}$} & \multirow{3}{*}{$d_{ij}=b\times\eta_{1i}\times\exp(-b\times{t_{j}})$} & $\eta_{0i}$: the initial status \\ & & $\eta_{1i}$: the change from initial status to asymptotic level \\ & & $\exp(-b)$: the ratio of rate-of-change at $t_{j}$ to that at $t_{j-1}$ \\
\hline
\hline
\multicolumn{3}{c}{\textbf{Jenss-Bayley function}}\\
\hline
\textbf{Growth Status} & \textbf{Growth Rate} & \textbf{Coef. related to developmental theory} \\
\hline
\multirow{4}{*}{$y_{ij}=\eta_{0i}+\eta_{1i}\times t_{j}+\eta_{2i}\times(\exp(c\times t_{j})-1)+\epsilon_{ij}$} & \multirow{4}{*}{$d_{ij}=\eta_{1i}+c\times\eta_{2i}\times\exp(c\times{t_{j}})$} & $\eta_{0i}$: the initial status \\ & & $\eta_{1i}$: the slope of linear asymptote \\ & & $\eta_{2i}$: the change from initial status to the intercept of linear asymptote \\ & & $\exp(c)$: the ratio of acceleration at $t_{j}$ to that at $t_{j-1}$ \\
\hline
\hline
\multicolumn{3}{c}{\textbf{Nonparametric function}}\\
\hline
\textbf{Growth Status} & \textbf{Growth Rate} & \textbf{Coef. related to developmental theory} \\
\hline
\multirow{4}{*}{$y_{ij}=\eta_{0i}+\eta_{1i}(1+\sum_{j=2}^{J-1}\lambda_{j})+\epsilon_{ij}$} & \multirow{4}{*}{$d_{ij}=a_{j}\times\eta_{1i}$} & $\eta_{0i}$: the initial status \\ & & $\eta_{1i}$: the change (LGCM) or the slope (LCSM) during the $1^{st}$ interval \\ & & $\lambda_{j}$: the quotient of the change that occurs in the $j^{th}$ interval to that of the $1^{st}$ interval \\ & & $a_{j}$: the quotient of the slope of the $j^{th}$ interval to that of the $1^{st}$ interval \\ 
\hline
\hline
\end{tabular}
\label{tbl:model_summary}
\end{threeparttable}}%
\end{table}

\begin{table}
\centering
\begin{threeparttable}
\caption{Performance Measures for Evaluating an Estimate ($\hat{\theta}$) of Parameter ($\theta$)}
\begin{tabular}{p{4cm}p{4.5cm}p{6cm}}
\hline
\hline
\textbf{Criteria} & \textbf{Definition} & \textbf{Estimate} \\
\hline
Relative Bias & $E_{\hat{\theta}}(\hat{\theta}-\theta)/\theta$ & $\sum_{s=1}^{S}(\hat{\theta}_{s}\tnote{a}-\theta)/\theta S\tnote{b}$ \\
Empirical SE & $\sqrt{Var(\hat{\theta})}$ & $\sqrt{\sum_{s=1}^{S}(\hat{\theta}_{s}-\bar{\theta}\tnote{c})^{2}/(S-1)}$ \\
Relative RMSE & $\sqrt{E_{\hat{\theta}}(\hat{\theta}-\theta)^{2}}/\theta$ & $\sqrt{\sum_{s=1}^{S}(\hat{\theta}_{s}-\theta)^{2}/S}/\theta$ \\
Coverage Probability & $Pr(\hat{\theta}_{\text{lower}}\le\theta\le\hat{\theta}_{\text{upper}})$ & $\sum_{s=1}^{S}I(\hat{\theta}_{\text{lower},s}\le\theta\le\hat{\theta}_{\text{upper},s})\tnote{d}/S$\\
\hline
\hline
\end{tabular}
\label{tbl:metric}
\begin{tablenotes}
\small
\item[a] {$\hat{\theta}_{s}$: the estimate of $\theta$ from the $s^{th}$ replication}\\ 
\item[b] {$S$: the number of replications and set as $1,000$ in our simulation study}\\
\item[c] {$\bar{\theta}$: the mean of $\hat{\theta}_{s}$'s across replications}\\
\item[d] {$I()$: an indicator function}
\end{tablenotes}
\end{threeparttable}
\end{table}

\begin{table}
\centering
\resizebox{1.15\textwidth}{!}{
\begin{threeparttable}
\setlength{\tabcolsep}{4pt}
\renewcommand{\arraystretch}{0.6}
\caption{Simulation Design for Each Latent Change Score Model in the Framework of Individual Measurement Occasions}
\begin{tabular}{p{6cm} p{13.6cm}}
\hline
\hline
\multicolumn{2}{c}{\textbf{Common Conditions across Four Models}}\\
\hline
\hline
Intercept & $\eta_{0i}\sim N(50, 5^{2})$ (i.e., $\mu_{\eta_{0}}=50$, $\psi_{00}=25$) \\
\hline
Correlations of Growth Factors & $\rho=0.3$ \\
\hline
\multirow{3}{*}{Time ($t_{j}$)} & $6$ equally-spaced: $t_{j}=0, 1.00, 2.00, 3.00, 4.00, 5.00$\\
& $10$ equally-spaced: $t_{j}=0, 1.00, 2.00, 3.00, 4.00, 5.00, 6.00, 7.00, 8.00, 9.00$\\
& $10$ unequally-spaced: $t_{j}=0, 0.75, 1.50, 2.25, 3.00, 3.75, 4.50, 6.00, 7.50, 9.00$\\
\hline
Individual $t_{ij}$ & $t_{ij} \sim U(t_{j}-\Delta, t_{j}+\Delta)$ ($\Delta=0.25$) \\
\hline
Sample Size & $n=200, 500$ \\
\hline
Residual Variance & $\theta_{\epsilon}=1, 2$ \\
\hline
\hline
\multicolumn{2}{c}{\textbf{Latent Change Score Model with Nonparametric Function}}\\
\hline
\hline
\textbf{Variables} & \textbf{Conditions} \\
\hline
Shape Factor & $\eta_{1i}\sim N(3.0, 1.0^{2})$ (i.e., $\mu_{\eta_{1}}=5$, $\psi_{11}=1$) \\
\hline
\multirow{4}{*}{Relative Rate-of-Change\tnote{a}} & $6$ waves: $\gamma_{1}=1.0$ (fixed), $\gamma_{2/3/4/5}=0.8/0.6/0.4/0.2$ \\
& $6$ waves: $\gamma_{1}=1.0$ (fixed), $\gamma_{2/3/4/5}=1.2/1.4/1.6/1.8$ \\
& $10$ waves: $\gamma_{1}=1.0$ (fixed), $\gamma_{2/3/4/5/6/7/8/9}=0.9/0.8/0.7/0.6/0.5/0.4/0.3/0.2$ \\
& $10$ waves: $\gamma_{1}=1.0$ (fixed), $\gamma_{2/3/4/5/6/7/8/9}=1.1/1.2/1.3/1.4/1.5/1.6/1.7/1.8$ \\
\hline
\hline
\multicolumn{2}{c}{\textbf{Latent Change Score Model with Quadratic Function}}\\
\hline
\hline
\textbf{Variables} & \textbf{Conditions} \\
\hline
\multirow{2}{*}{Linear Slope} & $6$ waves: $\eta_{1i}\sim N(16.0, 1.0^{2})$ (i.e., $\mu_{\eta_{1}}=16$, $\psi_{11}=1$) \\
& $10$ waves: $\eta_{1i}\sim N(20.0, 1.0^{2})$ (i.e., $\mu_{\eta_{1}}=20$, $\psi_{11}=1$) \\
\hline
\multirow{2}{*}{Quadratic Slope} & $6$ waves: $\eta_{2i}\sim N(-1.5, 0.3^{2})$ (i.e., $\mu_{\eta_{2}}=-1.5$, $\psi_{22}=0.09$) \\
& $10$ waves: $\eta_{2i}\sim N(-1.0, 0.3^{2})$ (i.e., $\mu_{\eta_{2}}=-1.0$, $\psi_{22}=0.09$) \\
\hline
\hline
\multicolumn{2}{c}{\textbf{Latent Change Score Model with Negative Exponential Function}}\\
\hline
\hline
Vertical Distance\tnote{b} & $\eta_{1i}\sim N(30, 3.0^{2})$ (i.e., $\mu_{\eta_{1}}=30.0$, $\psi_{11}=9.0$) \\
\hline
Log-ratio of Growth Rate & $b=0.4, 0.8$ \\
\hline
\hline
\multicolumn{2}{c}{\textbf{Latent Change Score Model with Jenss-Bayley Function}}\\
\hline
\hline
Vertical Distance\tnote{c} & $\eta_{2i}\sim N(-30, 3.0^{2})$ (i.e., $\mu_{\eta_{2}}=-30$, $\psi_{22}=25$) \\
\hline
\multirow{2}{*}{Slope of the Linear Asymptote}
& $\eta_{1i}\sim N(2.5, 1.0^{2})$ (i.e., $\mu_{\eta_{1}}=2.5$, $\psi_{11}=1.00$)\\
& $\eta_{1i}\sim N(1.0, 0.4^{2})$ (i.e., $\mu_{\eta_{1}}=1.0$, $\psi_{11}=0.16$)\\
\hline
Log-ratio of Growth Acceleration & $c=-0.7$ (i.e., $\exp(c)=0.5$) \\
\hline
\hline
\end{tabular}
\label{tbl:simu_design}
\begin{tablenotes}
\small
\item[a] {Relative rate-of-change is defined as the absolute rate-of-change over the shape factor.} \\
\item[b] {Vertical distance in the negative exponential function is the distance between the intercept (i.e., the initial status) and the asymptotic level.}\\
\item[c] {Vertical distance in the Jenss-Bayley function is the distance between the intercept (i.e., the initial status) and the intercept of the linear asymptote.}
\end{tablenotes}
\end{threeparttable}}
\end{table}

\begin{table}
\centering
\resizebox{1.0\textwidth}{!}{
\begin{threeparttable}
\small
\caption{Mean and Standard Deviation of Reading IRT Scores for the ECLS-K: 2011 Analytic Sample}
\begin{tabular}{llrr}
\hline
\hline
Grade & Semester & Mean (SD) of Observed Scores & Mean (SD) of Observed Change-from-baseline \\
\hline
\multirow{2}{*}{Kindergarten} & Fall & $54.63$ ($12.56$) & --- \\
& Spring & $69.10$ ($15.33$) & $14.47$ ($8.56$) \\
\multirow{2}{*}{First grade}  & Fall & $78.24$ ($17.57$) & $23.61$ ($11.21$) \\
& Spring & $95.85$ ($17.70$) & $41.23$ ($13.48$) \\
\multirow{2}{*}{Second grade} & Fall & $102.04$ ($17.60$) & $47.41$ ($13.73$) \\
& Spring & $112.52$ ($17.08$) & $57.90$ ($13.73$) \\
Third grade & Spring & $120.56$ ($15.06$) & $65.93$ ($13.40$) \\
Fourth grade & Spring & $129.18$ ($14.63$) & $74.55$ ($12.63$) \\
Fifth grade & Spring & $135.94$ ($14.89$) & $81.31$ ($13.90$) \\
\hline
\hline
\end{tabular}
\label{tbl:raw}
\begin{tablenotes}
\small
\item[] Note: IRT = item response theory; ECLS-K: 2011 = Early Childhood Longitudinal Study, Kindergarten Class 2010-11; SD = standard deviation.
\end{tablenotes}
\end{threeparttable}}
\end{table}

\begin{table}
\centering
\resizebox{1.0\textwidth}{!}{
\begin{threeparttable}
\small
\setlength{\tabcolsep}{4pt}
\renewcommand{\arraystretch}{0.6}
\caption{Summary of Model Fit Information For the Models}
\begin{tabular}{lrrrrr}
\hline
\hline
\textbf{Model} & \textbf{-2ll} & \textbf{AIC}  & \textbf{BIC}  & \textbf{\# of Para.} & \textbf{Residual}  \\
\hline
Latent Basis Growth Model & $26205.17$ & $26232.11$ & $26241.81$ & $13$ & $49.208$ \\
Quadratic Latent Change Score Model & $26527.88$ & $26548.45$ & $26556.07$ & $10$ & $48.753$ \\
Negative Exponential Latent Change Score Model & $26617.17$ & $26631.46$ & $26636.90$ & $7$ & $53.997$ \\
Jenss-Bayley Latent Change Score Model & $26392.76$ & $26415.44$ & $26423.76$ & $11$ & $45.749$ \\
\hline
Quadratic Latent Growth Curve Model & $26464.57$ & $26485.14$ & $26492.76$ & $10$ & $48.933$ \\
Negative Exponential Latent Growth Curve Model & $26609.86$ & $26624.14$ & $26629.59$ & $7$ & $54.015$ \\
Jenss-Bayley Latent Growth Curve Model & $26362.91$ & $26385.59$ & $26393.91$ & $11$ & $46.120$ \\
\hline
\hline
\end{tabular}
\label{tbl:info}
\end{threeparttable}}
\end{table}

\begin{table}
\centering
\resizebox{1.0\textwidth}{!}{
\begin{threeparttable}
\setlength{\tabcolsep}{4pt}
\renewcommand{\arraystretch}{0.6}
\caption{Estimates of Nonparametric Latent Change Score Model}
\begin{tabular}{crrcrrcrr}
\hline
\hline
\textbf{Para.} & \textbf{Estimate (SE)} & \textbf{P value} & \textbf{Para.} & \textbf{Estimate (SE)} & \textbf{P value} & \textbf{Para.} & \textbf{Estimate (SE)} & \textbf{P value}\\
\hline
$\mu_{\eta_{0}}$ & $54.724$ ($0.804$) & $<0.0001^{\ast}$\tnote{a} & & & & & &\\
$\mu_{\eta_{1}}$ & $29.401$ ($1.015$) & $<0.0001^{\ast}$ & & & & & &\\
$\psi_{00}$ & $210.702$ ($16.431$) & $<0.0001^{\ast}$ & & & & & &\\
$\psi_{01}$ & $-22.422$ ($4.318$) & $<0.0001^{\ast}$ & & & & & &\\
$\psi_{11}$ & $19.562$ ($2.262$) & $<0.0001^{\ast}$ & & & & & & \\
$\theta_{\epsilon}$ & $49.208$ ($1.315$) & $<0.0001^{\ast}$ & & & & & &\\
\hline
$\gamma_{1}$\tnote{b} & $1.000$ ($0.000$) & --- & $\mu_{d_{1}}$\tnote{c} & $29.401$ ($1.015$) & $<0.0001^{\ast}$ & $\phi_{d_{1}}$\tnote{d} & $19.562$ ($2.272$) & $<0.0001^{\ast}$ \\
$\gamma_{2}$\tnote{b} & $0.641$ ($0.047$) & $<0.0001^{\ast}$ & $\mu_{d_{2}}$\tnote{c} & $18.851$ ($0.962$) & $<0.0001^{\ast}$ & $\phi_{d_{2}}$\tnote{d} & $8.042$ ($1.113$) & $<0.0001^{\ast}$ \\
$\gamma_{3}$\tnote{b} & $1.159$ ($0.051$) & $<0.0001^{\ast}$ & $\mu_{d_{3}}$\tnote{c} & $34.074$ ($0.962$) & $<0.0001^{\ast}$ & $\phi_{d_{3}}$\tnote{d} & $26.276$ ($2.927$) & $<0.0001^{\ast}$ \\
$\gamma_{4}$\tnote{b} & $0.472$ ($0.037$) & $<0.0001^{\ast}$ & $\mu_{d_{4}}$\tnote{c} & $13.882$ ($0.995$) & $<0.0001^{\ast}$ & $\phi_{d_{4}}$\tnote{d} & $4.361$ ($0.742$) & $<0.0001^{\ast}$ \\
$\gamma_{5}$\tnote{b} & $0.660$ ($0.038$) & $<0.0001^{\ast}$ & $\mu_{d_{5}}$\tnote{c} & $19.416$ ($0.917$) & $<0.0001^{\ast}$ & $\phi_{d_{5}}$\tnote{d} & $8.532$ ($1.150$) & $<0.0001^{\ast}$ \\
$\gamma_{6}$\tnote{b} & $0.286$ ($0.020$) & $<0.0001^{\ast}$ & $\mu_{d_{6}}$\tnote{c} & $8.411$ ($0.506$) & $<0.0001^{\ast}$ & $\phi_{d_{6}}$\tnote{d} & $1.601$ ($0.243$) & $<0.0001^{\ast}$ \\
$\gamma_{7}$\tnote{b} & $0.283$ ($0.019$) & $<0.0001^{\ast}$ & $\mu_{d_{7}}$\tnote{c} & $8.325$ ($0.490$) & $<0.0001^{\ast}$ & $\phi_{d_{7}}$\tnote{d} & $1.568$ ($0.236$) & $<0.0001^{\ast}$ \\
$\gamma_{8}$\tnote{b} & $0.231$ ($0.018$) & $<0.0001^{\ast}$ & $\mu_{d_{8}}$\tnote{c} & $6.796$ ($0.491$) & $<0.0001^{\ast}$ & $\phi_{d_{8}}$\tnote{d} & $1.045$ ($0.180$) & $<0.0001^{\ast}$ \\
\hline
\hline
\end{tabular}
\label{tbl:est_LBGM}
\begin{tablenotes}
\small
\item[a] $^{\ast}$ indicates statistical significance at $0.05$ level. \\
\item[b] $\gamma_{1}=1$, which is fixed in our model specification. $\gamma_{j}$ ($j=2,\dots,8$) indicates the relative rate-of-change (i.e., the absolute rate-of-change over the shape factor) in the $j^{th}$ time interval.\\
\item[c] $\mu_{d_{j}}=\mu_{\eta_{1}}\gamma_{j}$ ($j=1,\dots,8$) indicates the estimated mean value of the absolute rate-of-change in the $j^{th}$ time interval.\\
\item[d] $\phi_{d_{j\_\text{mid}}}$ ($j=1,\dots,8$) indicates the estimated variance of the absolute rate-of-change the $j^{th}$ interval. \\
\end{tablenotes}
\end{threeparttable}}
\end{table}

\begin{table}
\centering
\resizebox{1.0\textwidth}{!}{
\begin{threeparttable}
\setlength{\tabcolsep}{4pt}
\renewcommand{\arraystretch}{0.6}
\caption{Estimates of Quadratic Latent Change Score Model}
\begin{tabular}{crrcrrcrr}
\hline
\hline
\textbf{Para.} & \textbf{Estimate (SE)} & \textbf{P value} & \textbf{Para.} & \textbf{Estimate (SE)} & \textbf{P value} & \textbf{Para.} & \textbf{Estimate (SE)} & \textbf{P value}\\
\hline
$\mu_{\eta_{0}}$ & $55.883$ ($0.713$) & $<0.0001^{\ast}$\tnote{a} & $\mu_{d_{1\_\text{mid}}}$\tnote{b} & $26.871$ ($0.360$) & $<0.0001^{\ast}$ & $\phi_{d_{1\_\text{mid}}}$\tnote{c} & $32.092$ ($3.750$) & $<0.0001^{\ast}$ \\
$\mu_{\eta_{1}}$ & $31.023$ ($0.455$) & $<0.0001^{\ast}$ & $\mu_{d_{2\_\text{mid}}}$\tnote{b} & $24.444$ ($0.306$) & $<0.0001^{\ast}$ & $\phi_{d_{2\_\text{mid}}}$\tnote{c} & $24.021$ ($2.719$) & $<0.0001^{\ast}$ \\
$\mu_{\eta_{2}}$ & $-2.489$ ($0.062$) & $<0.0001^{\ast}$ & $\mu_{d_{3\_\text{mid}}}$\tnote{b} & $21.942$ ($0.253$) & $<0.0001^{\ast}$ & $\phi_{d_{3\_\text{mid}}}$\tnote{c} & $17.285$ ($1.871$) & $<0.0001^{\ast}$ \\
$\psi_{00}$ & $175.965$ ($14.393$) & $<0.0001^{\ast}$ & $\mu_{d_{4\_\text{mid}}}$\tnote{b} & $19.478$ ($0.206$) & $<0.0001^{\ast}$ & $\phi_{d_{4\_\text{mid}}}$\tnote{c} & $12.217$ ($1.246$) & $<0.0001^{\ast}$ \\
$\psi_{01}$ & $9.258$ ($7.306$) & $0.2050$ & $\mu_{d_{5\_\text{mid}}}$\tnote{b} & $16.968$ ($0.168$) & $<0.0001^{\ast}$ & $\phi_{d_{5\_\text{mid}}}$\tnote{c} & $8.655$ ($0.825$) & $<0.0001^{\ast}$ \\
$\psi_{02}$ & $-2.860$ ($0.965$) & $0.0031^{\ast}$ & $\mu_{d_{6\_\text{mid}}}$\tnote{b} & $13.246$ ($0.147$) & $<0.0001^{\ast}$ & $\phi_{d_{6\_\text{mid}}}$\tnote{c} & $6.343$ ($0.608$) & $<0.0001^{\ast}$ \\
$\psi_{11}$ & $49.393$ ($5.992$) & $<0.0001^{\ast}$ & $\mu_{d_{7\_\text{mid}}}$\tnote{b} & $8.355$ ($0.196$) & $<0.0001^{\ast}$ & $\phi_{d_{7\_\text{mid}}}$\tnote{c} & $8.703$ ($1.081$) & $<0.0001^{\ast}$ \\
$\psi_{12}$ & $-5.848$ ($0.780$) & $<0.0001^{\ast}$ & $\mu_{d_{8\_\text{mid}}}$\tnote{b} & $3.378$ ($0.293$) & $<0.0001^{\ast}$ & $\phi_{d_{8\_\text{mid}}}$\tnote{c} & $17.399$ ($2.415$) & $<0.0001^{\ast}$ \\
$\psi_{22}$ & $0.794$ ($0.108$) & $<0.0001^{\ast}$ & & & & & & \\
$\theta_{\epsilon}$ & $48.753$ ($1.413$) & $<0.0001^{\ast}$ & & & & & & \\
\hline
\hline
\end{tabular}
\label{tbl:est_QUAD}
\begin{tablenotes}
\small
\item[a] $^{\ast}$ indicates statistical significance at $0.05$ level. \\
\item[b] $\mu_{d_{j\_\text{mid}}}$ ($j=1,\dots,8$) indicates the estimated mean value of the instantaneous slope midway the $j^{th}$ interval. \\
\item[c] $\phi_{d_{j\_\text{mid}}}$ ($j=1,\dots,8$) indicates the estimated variance of the instantaneous slope midway the $j^{th}$ interval. \\
\end{tablenotes}
\end{threeparttable}}
\end{table}

\begin{table}
\centering
\resizebox{1.0\textwidth}{!}{
\begin{threeparttable}
\setlength{\tabcolsep}{4pt}
\renewcommand{\arraystretch}{0.6}
\caption{Estimates of Negative Exponential Latent Change Score Model}
\begin{tabular}{crrcrrcrr}
\hline
\hline
\textbf{Para.} & \textbf{Estimate (SE)} & \textbf{P value} & \textbf{Para.} & \textbf{Estimate (SE)} & \textbf{P value} & \textbf{Para.} & \textbf{Estimate (SE)} & \textbf{P value}\\
\hline
$\mu_{\eta_{0}}$ & $54.417$ ($0.785$) & $<0.0001^{\ast}$\tnote{a} & $\mu_{d_{1\_\text{mid}}}$\tnote{b} & $30.669$ ($0.450$) & $<0.0001^{\ast}$ & $\phi_{d_{1\_\text{mid}}}$\tnote{c} & $28.748$ ($2.890$) & $<0.0001^{\ast}$ \\
$\mu_{\eta_{1}}$ & $118.531$ ($1.211$) & $<0.0001^{\ast}$ & $\mu_{d_{2\_\text{mid}}}$\tnote{b} & $25.919$ ($0.326$) & $<0.0001^{\ast}$ & $\phi_{d_{2\_\text{mid}}}$\tnote{c} & $20.533$ ($2.019$) & $<0.0001^{\ast}$ \\
$\psi_{00}$ & $209.538$ ($16.562$) & $<0.0001^{\ast}$ & $\mu_{d_{3\_\text{mid}}}$\tnote{b} & $21.794$ ($0.238$) & $<0.0001^{\ast}$ & $\phi_{d_{3\_\text{mid}}}$\tnote{c} & $14.517$ ($1.401$) & $<0.0001^{\ast}$ \\
$\psi_{01}$ & $-47.027$ ($18.854$) & $0.0126^{\ast}$ & $\mu_{d_{4\_\text{mid}}}$\tnote{b} & $18.372$ ($0.184$) & $<0.0001^{\ast}$ & $\phi_{d_{4\_\text{mid}}}$\tnote{c} & $10.317$ ($0.980$) & $<0.0001^{\ast}$ \\
$\psi_{11}$ & $429.413$ ($40.461$) & $<0.0001^{\ast}$ & $\mu_{d_{5\_\text{mid}}}$\tnote{b} & $15.440$ ($0.156$) & $<0.0001^{\ast}$ & $\phi_{d_{5\_\text{mid}}}$\tnote{c} & $7.286$ ($0.684$) & $<0.0001^{\ast}$ \\
$b$ & $0.345$ ($0.006$) & $<0.0001^{\ast}$ & $\mu_{d_{6\_\text{mid}}}$\tnote{b} & $11.929$ ($0.143$) & $<0.0001^{\ast}$ & $\phi_{d_{6\_\text{mid}}}$\tnote{c} & $4.349$ ($0.405$) & $<0.0001^{\ast}$ \\
$\theta_{\epsilon}$ & $53.997$ ($1.452$) & $<0.0001^{\ast}$ & $\mu_{d_{7\_\text{mid}}}$\tnote{b} & $8.500$ ($0.138$) & $<0.0001^{\ast}$ & $\phi_{d_{7\_\text{mid}}}$\tnote{c} & $2.208$ ($0.207$) & $<0.0001^{\ast}$ \\
& & & $\mu_{d_{8\_\text{mid}}}$\tnote{b} & $6.020$ ($0.130$) & $<0.0001^{\ast}$ & $\phi_{d_{8\_\text{mid}}}$\tnote{c} & $1.108$ ($0.106$) & $<0.0001^{\ast}$ \\
\hline
\hline
\end{tabular}
\label{tbl:est_EXP}
\begin{tablenotes}
\small
\item[a] $^{\ast}$ indicates statistical significance at $0.05$ level. \\
\item[b] $\mu_{d_{j\_\text{mid}}}$ ($j=1,\dots,8$) indicates the estimated mean value of the instantaneous slope midway the $j^{th}$ interval. \\
\item[c] $\phi_{d_{j\_\text{mid}}}$ ($j=1,\dots,8$) indicates the estimated variance of the instantaneous slope midway the $j^{th}$ interval. \\
\end{tablenotes}
\end{threeparttable}}
\end{table}

\begin{table}
\centering
\resizebox{1.0\textwidth}{!}{
\begin{threeparttable}
\setlength{\tabcolsep}{4pt}
\renewcommand{\arraystretch}{0.6}
\caption{Estimates of Jenss-Bayley Latent Change Score Model}
\begin{tabular}{crrcrrcrr}
\hline
\hline
\textbf{Para.} & \textbf{Estimate (SE)} & \textbf{P value} & \textbf{Para.} & \textbf{Estimate (SE)} & \textbf{P value} & \textbf{Para.} & \textbf{Estimate (SE)} & \textbf{P value}\\
\hline
$\mu_{\eta_{0}}$ & $54.340$ ($0.707$) & $<0.0001^{\ast}$\tnote{a} & $\mu_{d_{1\_\text{mid}}}$\tnote{b} & $30.775$ ($0.571$) & $<0.0001^{\ast}$ & $\phi_{d_{1\_\text{mid}}}$\tnote{c} & $58.575$ ($7.023$) & $<0.0001^{\ast}$ \\
$\mu_{\eta_{1}}$ & $-1.458$ ($1.278$) & $0.2541$ & $\mu_{d_{2\_\text{mid}}}$\tnote{b} & $26.138$ ($0.379$) & $<0.0001^{\ast}$ & $\phi_{d_{2\_\text{mid}}}$\tnote{c} & $35.740$ ($3.953$) & $<0.0001^{\ast}$ \\
$\mu_{\eta_{2}}$ & $-131.988$ ($11.755$) & $<0.0001^{\ast}$ & $\mu_{d_{3\_\text{mid}}}$\tnote{b} & $22.056$ ($0.270$) & $<0.0001^{\ast}$ & $\phi_{d_{3\_\text{mid}}}$\tnote{c} & $21.016$ ($2.153$) & $<0.0001^{\ast}$ \\
$\psi_{00}$ & $167.39$ ($13.886$) & $<0.0001^{\ast}$ & $\mu_{d_{4\_\text{mid}}}$\tnote{b} & $18.626$ ($0.220$) & $<0.0001^{\ast}$ & $\phi_{d_{4\_\text{mid}}}$\tnote{c} & $12.531$ ($1.202$) & $<0.0001^{\ast}$ \\
$\psi_{01}$ & $-32.172$ ($5.716$) & $<0.0001^{\ast}$ & $\mu_{d_{5\_\text{mid}}}$\tnote{b} & $15.647$ ($0.193$) & $<0.0001^{\ast}$ & $\phi_{d_{5\_\text{mid}}}$\tnote{c} & $8.045$ ($0.736$) & $<0.0001^{\ast}$ \\
$\psi_{02}$ & $-186.537$ ($48.621$) & $0.0001^{\ast}$ & $\mu_{d_{6\_\text{mid}}}$\tnote{b} & $12.022$ ($0.168$) & $<0.0001^{\ast}$ & $\phi_{d_{6\_\text{mid}}}$\tnote{c} & $6.201$ ($0.618$) & $<0.0001^{\ast}$ \\
$\psi_{11}$ & $34.157$ ($6.445$) & $<0.0001^{\ast}$ & $\mu_{d_{7\_\text{mid}}}$\tnote{b} & $8.400$ ($0.189$) & $<0.0001^{\ast}$ & $\phi_{d_{7\_\text{mid}}}$\tnote{c} & $8.321$ ($1.012$) & $<0.0001^{\ast}$ \\
$\psi_{12}$ & $271.19$ ($56.553$) & $<0.0001^{\ast}$ & $\mu_{d_{8\_\text{mid}}}$\tnote{b} & $5.711$ ($0.291$) & $<0.0001^{\ast}$ & $\phi_{d_{8\_\text{mid}}}$\tnote{c} & $12.458$ ($1.644$) & $<0.0001^{\ast}$ \\
$\psi_{22}$ & $2630.396$ ($516.047$) & $<0.0001^{\ast}$ & & & & & & \\
$c$ & $-0.318$ ($0.025$) & $<0.0001^{\ast}$ & & & & & & \\
$\theta_{\epsilon}$ & $45.749$ ($1.327$) &  $<0.0001^{\ast}$ & & & & & & \\
\hline
\hline
\end{tabular}
\label{tbl:est_JB}
\begin{tablenotes}
\small
\item[a] $^{\ast}$ indicates statistical significance at $0.05$ level. \\
\item[b] $\mu_{d_{j\_\text{mid}}}$ ($j=1,\dots,8$) indicates the estimated mean value of the instantaneous slope midway the $j^{th}$ interval. \\
\item[c] $\phi_{d_{j\_\text{mid}}}$ ($j=1,\dots,8$) indicates the estimated variance of the instantaneous slope midway the $j^{th}$ interval. \\
\end{tablenotes}
\end{threeparttable}}
\end{table}

\setcounter{table}{0}
\renewcommand{\thetable}{C\arabic{table}}

\begin{table}
\centering
\resizebox{1.0\textwidth}{!}{
\begin{threeparttable}
\setlength{\tabcolsep}{4pt}
\renewcommand{\arraystretch}{0.6}
\caption{Summary of Performance Metrics of Nonparametric Latent Change Score Model (Latent Basis Growth Model)}
\begin{tabular}{lrrrr}
\hline
\hline
\textbf{Para.} & \textbf{Relative Bias} & \textbf{Empirical SE\tnote{a}} & \textbf{Relative RMSE\tnote{b}} & \textbf{Coverage Probability} \\
\hline
$\mu_{\eta_{0}}$ & $0.0000$ ($-0.0003$, $0.0006$) & $0.2958$ ($0.2209$, $0.3847$) & $0.0059$ ($0.0044$, $0.0077$) & $0.9460$ ($0.9370$, $0.9590$) \\
$\mu_{\eta_{1}}$ & $0.0003$ ($-0.0031$, $0.0031$) & $0.1168$ ($0.0721$, $0.1873$) & $0.0390$ ($0.0240$, $0.0625$) & $0.9510$ ($0.9370$, $0.9610$) \\
$\psi_{00}$ & $-0.0039$ ($-0.0094$, $0.0005$) & $2.0592$ ($1.5859$, $2.6213$) & $0.0824$ ($0.0634$, $0.1049$) & $0.9455$ ($0.9310$, $0.9580$) \\
$\psi_{01}$ & $-0.0073$ ($-0.0335$, $0.0090$) & $0.3159$ ($0.2296$, $0.4267$) & $0.2107$ ($0.1533$, $0.2863$) & $0.9490$ ($0.9370$, $0.9580$) \\
$\psi_{11}$ & $-0.0033$ ($-0.0123$, $0.0020$) & $0.1100$ ($0.0735$, $0.1603$) & $0.1099$ ($0.0741$, $0.1604$) & $0.9445$ ($0.9270$, $0.9560$) \\
$\gamma_{2}$\tnote{c} & $0.0014$ ($-0.0021$, $0.0069$) & $0.0543$ ($0.0316$, $0.0985$) & $0.0545$ ($0.0310$, $0.1019$) & $0.9495$ ($0.9360$, $0.9660$) \\
$\gamma_{3}$\tnote{c} & $0.0010$ ($-0.0013$, $0.0078$) & $0.0479$ ($0.0224$, $0.0872$) & $0.0492$ ($0.0238$, $0.0958$) & $0.9485$ ($0.9300$, $0.9610$) \\
$\gamma_{4}$\tnote{c} & $0.0008$ ($-0.0029$, $0.0049$) & $0.0469$ ($0.0200$, $0.0900$) & $0.0502$ ($0.0229$, $0.1180$) & $0.9530$ ($0.9400$, $0.9600$) \\
$\gamma_{5}$\tnote{c} & $0.0003$ ($-0.0034$, $0.0151$) & $0.0458$ ($0.0200$, $0.0938$) & $0.0520$ ($0.0218$, $0.2156$) & $0.9500$ ($0.9360$, $0.9660$) \\
$\gamma_{6}$\tnote{c,d} & $0.0014$ ($-0.0014$, $0.0058$) & $0.0469$ ($0.0200$, $0.1058$) & $0.0530$ ($0.0236$, $0.1221$) & $0.9455$ ($0.9360$, $0.9600$) \\
$\gamma_{7}$\tnote{c,d} & $0.0007$ ($-0.0033$, $0.0079$) & $0.0393$ ($0.0173$, $0.0949$) & $0.0510$ ($0.0236$, $0.1106$) & $0.9510$ ($0.9410$, $0.9600$) \\
$\gamma_{8}$\tnote{c,d} & $0.0003$ ($-0.0088$, $0.0070$) & $0.0406$ ($0.0141$, $0.1025$) & $0.0556$ ($0.0231$, $0.1427$) & $0.9505$ ($0.9390$, $0.9580$) \\
$\gamma_{9}$\tnote{c,d} & $0.0017$ ($-0.0012$, $0.0090$) & $0.0418$ ($0.0141$, $0.1058$) & $0.0653$ ($0.0224$, $0.2153$) & $0.9505$ ($0.9380$, $0.9560$) \\
$\theta_{\epsilon}$ & $-0.0032$ ($-0.0064$, $-0.0004$) & $0.0442$ ($0.0224$, $0.1005$) & $0.0332$ ($0.0213$, $0.0524$) & $0.9490$ ($0.9330$, $0.9610$) \\
\hline
\hline
\end{tabular}
\label{tbl:Metric_LBGM}
\begin{tablenotes}
\small
\item[a] SE: standard error.\\
\item[b] RMSE: root mean square error.\\
\item[c] $\gamma_{j}$ ($j=2,\dots,9$) indicates the relative rate-of-change (i.e., the absolute rate-of-change over the shape factor) in the $j^{th}$ time interval.\\
\item[d] The summary of metrics for $\gamma_{j}$ ($j=6,\dots,9$) was based on the estimates from the conditions with ten measurement occasions.
\end{tablenotes}
\end{threeparttable}}
\end{table}

\begin{table}
\centering
\resizebox{1.0\textwidth}{!}{
\begin{threeparttable}
\setlength{\tabcolsep}{4pt}
\renewcommand{\arraystretch}{0.6}
\caption{Summary of Performance Metrics of Quadratic Latent Change Score Model and Corresponding Latent Growth Curve Model}
\begin{tabular}{lrrrr}
\hline
\hline
\textbf{Para.} & \textbf{Relative Bias} & \textbf{Empirical SE\tnote{a}} & \textbf{Relative RMSE\tnote{b}} & \textbf{Coverage Probability} \\
\hline
\multicolumn{5}{c}{\textbf{Latent Change Score Model}} \\
\hline
$\mu_{\eta_{0}}$ & $0.0000$ ($-0.0001$, $0.0006$) & $0.2955$ ($0.2209$, $0.3726$) & $0.0059$ ($0.0044$, $0.0074$) & $0.9480$ ($0.9340$, $0.9580$) \\
$\mu_{\eta_{1}}$ & $0.0000$ ($-0.0003$, $0.0003$) & $0.0748$ ($0.0480$, $0.1109$) & $0.0040$ ($0.0024$, $0.0069$) & $0.9445$ ($0.9370$, $0.9530$) \\
$\mu_{\eta_{2}}$ & $0.0000$ ($-0.0008$, $0.0006$) & $0.0198$ ($0.0141$, $0.0265$) & $-0.0152$ ($-0.0216$, $-0.0099$) & $0.9490$ ($0.9370$, $0.9590$) \\
$\psi_{00}$ & $-0.0039$ ($-0.0091$, $0.0031$) & $2.1461$ ($1.5469$, $2.6927$) & $0.0858$ ($0.0619$, $0.1077$) & $0.9410$ ($0.9320$, $0.9610$)\tnote{c} \\
$\psi_{01}$ & $-0.0043$ ($-0.0110$, $0.0055$) & $0.3837$ ($0.2410$, $0.5782$) & $0.2558$ ($0.1610$, $0.3853$) & $0.9480$ ($0.9400$, $0.9630$) \\
$\psi_{02}$ & $-0.0060$ ($-0.0193$, $0.0149$) & $0.1013$ ($0.0707$, $0.1432$) & $0.2255$ ($0.1571$, $0.3184$) & $0.9460$ ($0.9310$, $0.9560$) \\
$\psi_{11}$ & $-0.0047$ ($-0.0136$, $-0.0028$) & $0.1179$ ($0.0714$, $0.2458$) & $0.1179$ ($0.0712$, $0.2458$) & $0.9430$ ($0.9340$, $0.9540$) \\
$\psi_{12}$ & $-0.0043$ ($-0.0180$, $0.0226$) & $0.0245$ ($0.0141$, $0.0469$) & $0.2629$ ($0.1638$, $0.5228$) & $0.9490$ ($0.9370$, $0.9670$) \\
$\psi_{22}$ & $-0.0042$ ($-0.0128$, $0.0023$) & $0.0100$ ($0.0000$, $0.0141$) & $0.1016$ ($0.0632$, $0.1695$) & $0.9475$ ($0.9240$, $0.9540$) \\
$\theta_{\epsilon}$ & $0.0000$ ($-0.0015$, $0.0022$) & $0.0485$ ($0.0245$, $0.1136$) & $0.0369$ ($0.0237$, $0.0579$) & $0.9515$ ($0.9420$, $0.9590$) \\
\hline
\hline
\multicolumn{5}{c}{\textbf{Latent Growth Curve Model}} \\
\hline
$\mu_{\eta_{0}}$ & $0.0000$ ($-0.0001$, $0.0006$) & $0.2955$ ($0.2209$, $0.3726$) & $0.0059$ ($0.0044$, $0.0074$) & $0.9480$ ($0.9340$, $0.9580$) \\
$\mu_{\eta_{1}}$ & $0.0000$ ($-0.0003$, $0.0003$) & $0.0748$ ($0.0480$, $0.1109$) & $0.0040$ ($0.0024$, $0.0069$) & $0.9445$ ($0.9370$, $0.9530$) \\
$\mu_{\eta_{2}}$ & $0.0000$ ($-0.0008$, $0.0006$) & $0.0198$ ($0.0141$, $0.0265$) & $-0.0152$ ($-0.0216$, $-0.0099$) & $0.9490$ ($0.9370$, $0.9590$) \\
$\psi_{00}$ & $-0.0039$ ($-0.0091$, $0.0031$) & $2.1461$ ($1.5469$, $2.6927$) & $0.0858$ ($0.0619$, $0.1077$) & $0.9410$ ($0.9330$, $0.9610$)\tnote{c} \\
$\psi_{01}$ & $-0.0043$ ($-0.0110$, $0.0055$) & $0.3837$ ($0.2410$, $0.5782$) & $0.2558$ ($0.1610$, $0.3853$) & $0.9480$ ($0.9400$, $0.9630$) \\
$\psi_{02}$ & $-0.0060$ ($-0.0193$, $0.0149$) & $0.1013$ ($0.0707$, $0.1432$) & $0.2255$ ($0.1571$, $0.3184$) & $0.9460$ ($0.9310$, $0.9560$) \\
$\psi_{11}$ & $-0.0047$ ($-0.0136$, $-0.0028$) & $0.1179$ ($0.0714$, $0.2458$) & $0.1179$ ($0.0712$, $0.2458$) & $0.9430$ ($0.9340$, $0.9540$) \\
$\psi_{12}$ & $-0.0043$ ($-0.0180$, $0.0226$) & $0.0245$ ($0.0141$, $0.0469$) & $0.2629$ ($0.1638$, $0.5228$) & $0.9490$ ($0.9370$, $0.9670$) \\
$\psi_{22}$ & $-0.0042$ ($-0.0128$, $0.0023$) & $0.0100$ ($0.0000$, $0.0141$) & $0.1016$ ($0.0632$, $0.1695$) & $0.9475$ ($0.9240$, $0.9540$) \\
$\theta_{\epsilon}$ & $0.0000$ ($-0.0015$, $0.0022$) & $0.0485$ ($0.0245$, $0.1136$) & $0.0369$ ($0.0237$, $0.0579$) & $0.9515$ ($0.9420$, $0.9590$) \\
\hline
\hline
\end{tabular}
\label{tbl:Metric_QUAD}
\begin{tablenotes}
\small
\item[a] SE: standard error.\\
\item[b] RMSE: root mean square error.\\
\item[c] These cells include a different summary of the performance metric of the quadratic latent change score model and that of the corresponding latent growth curve model.
\end{tablenotes}
\end{threeparttable}}
\end{table}

\begin{table}
\centering
\resizebox{1.0\textwidth}{!}{
\begin{threeparttable}
\setlength{\tabcolsep}{4pt}
\renewcommand{\arraystretch}{0.6}
\caption{Summary of Performance Metrics of Negative Exponential Latent Change Score Model and Corresponding Latent Growth Curve Model}
\begin{tabular}{lrrrr}
\hline
\hline
\textbf{Para.} & \textbf{Relative Bias} & \textbf{Empirical SE\tnote{a}} & \textbf{Relative RMSE\tnote{b}} & \textbf{Coverage Probability} \\
\hline
\multicolumn{5}{c}{\textbf{Latent Change Score Model}} \\
\hline
$\mu_{\eta_{0}}$ & $0.0000$ ($-0.0004$, $0.0005$) & $0.2963$ ($0.2198$, $0.3735$) & $0.0060$ ($0.0044$, $0.0075$) & $0.9500$ ($0.9280$, $0.9620$) \\
$\mu_{\eta_{1}}$ & $0.0130$ ($0.0059$, $0.0296$) & $0.2043$ ($0.1382$, $0.3058$) & $0.0159$ ($0.0077$, $0.0307$) & $0.5285$ ($0.0000$, $0.8910$) \\
$b$ & $-0.0015$ ($-0.0036$, $0.0005$) & $0.0000$ ($0.0000$, $0.0100$) & $0.0069$ ($0.0041$, $0.0177$) & $0.9395$ ($0.8610$, $0.9590$) \\
$\psi_{00}$ & $-0.0029$ ($-0.0076$, $0.0023$) & $2.1559$ ($1.6138$, $2.6852$) & $0.0863$ ($0.0645$, $0.1074$) & $0.9435$ ($0.9270$, $0.9540$) \\
$\psi_{01}$ & $0.0102$ ($-0.0037$, $0.0317$) & $1.0082$ ($0.7222$, $1.3685$) & $0.2248$ ($0.1623$, $0.3041$) & $0.9530$ ($0.9370$, $0.9670$) \\
$\psi_{11}$ & $0.0226$ ($0.0029$, $0.0636$) & $0.9249$ ($0.6204$, $1.3684$) & $0.1086$ ($0.0696$, $0.1571$) & $0.9485$ ($0.8910$, $0.9650$) \\
$\theta_{\epsilon}$ & $0.0000$ ($-0.0025$, $0.0013$) & $0.0458$ ($0.0224$, $0.0990$) & $0.0336$ ($0.0214$, $0.0506$) & $0.9485$ ($0.9360$, $0.9630$) \\
\hline
\hline
\multicolumn{5}{c}{\textbf{Latent Growth Curve Model}} \\
\hline
$\mu_{\eta_{0}}$ & $0.0000$ ($-0.0004$, $0.0005$) & $0.2963$ ($0.2198$, $0.3735$) & $0.0060$ ($0.0044$, $0.0075$) & $0.9495$ ($0.9300$, $0.9620$) \\
$\mu_{\eta_{1}}$ & $0.0000$ ($-0.0003$, $0.0003$) & $0.2043$ ($0.1353$, $0.3077$) & $0.0068$ ($0.0045$, $0.0103$) & $0.9480$ ($0.9330$, $0.9620$) \\
$b$ & $0.0000$ ($-0.0004$, $0.0005$) & $0.0000$ ($0.0000$, $0.0100$) & $0.0063$ ($0.0035$, $0.0177$) & $0.9530$ ($0.9240$, $0.9600$) \\
$\psi_{00}$ & $-0.0029$ ($-0.0076$, $0.0023$) & $2.1559$ ($1.6137$, $2.6853$) & $0.0863$ ($0.0645$, $0.1074$) & $0.9435$ ($0.9260$, $0.9540$) \\
$\psi_{01}$ & $-0.0023$ ($-0.0106$, $0.0041$) & $0.9960$ ($0.7010$, $1.3585$) & $0.2212$ ($0.1557$, $0.3017$) & $0.9515$ ($0.9370$, $0.9620$) \\
$\psi_{11}$ & $-0.0032$ ($-0.0093$, $0.0034$) & $0.9125$ ($0.6132$, $1.3484$) & $0.1016$ ($0.0681$, $0.1497$) & $0.9435$ ($0.9290$, $0.9610$) \\
$\theta_{\epsilon}$ & $-0.0007$ ($-0.0026$, $0.0013$) & $0.0458$ ($0.0224$, $0.0990$) & $0.0335$ ($0.0214$, $0.0507$) & $0.9485$ ($0.9370$, $0.9630$) \\
\hline
\hline
\end{tabular}
\label{tbl:Metric_EXP}
\begin{tablenotes}
\small
\item[a] SE: standard error.\\
\item[b] RMSE: root mean square error.\\
\end{tablenotes}
\end{threeparttable}}
\end{table}

\begin{table}
\centering
\resizebox{1.0\textwidth}{!}{
\begin{threeparttable}
\setlength{\tabcolsep}{4pt}
\renewcommand{\arraystretch}{0.6}
\caption{Summary of Performance Metrics of Jenss-Bayley Latent Change Score Model and Corresponding Latent Growth Curve Model}
\begin{tabular}{lrrrr}
\hline
\hline
\textbf{Para.} & \textbf{Relative Bias} & \textbf{Empirical SE\tnote{a}} & \textbf{Relative RMSE\tnote{b}} & \textbf{Coverage Probability} \\
\hline
\multicolumn{5}{c}{\textbf{Latent Change Score Model}} \\
\hline
$\mu_{\eta_{0}}$ & $-0.0002$ ($-0.0004$, $0.0002$) & $0.2919$ ($0.2216$, $0.3686$) & $0.0058$ ($0.0044$, $0.0074$) & $0.9495$ ($0.9350$, $0.9580$) \\
$\mu_{\eta_{1}}$ & $-0.0021$ ($-0.0299$, $0.0091$) & $0.0624$ ($0.0245$, $0.1947$) & $0.0346$ ($0.0193$, $0.1930$) & $0.9470$ ($0.9270$, $0.9660$) \\
$\mu_{\eta_{2}}$ & $0.0232$ ($0.0125$, $0.0289$) & $0.3126$ ($0.1778$, $1.0191$) & $-0.0248$ ($-0.0443$, $-0.0138$) & $0.5930$ ($0.0180$, $0.9100$) \\
$c$ & $-0.0019$ ($-0.0065$, $0.0002$) & $0.0100$ ($0.0000$, $0.0265$) & $-0.0137$ ($-0.0384$, $-0.0069$) & $0.9435$ ($0.9310$, $0.9590$) \\
$\psi_{00}$ & $-0.0033$ ($-0.0097$, $0.0012$) & $2.1538$ ($1.5806$, $2.7194$) & $0.0864$ ($0.0632$, $0.1091$) & $0.9410$ ($0.9260$, $0.9600$) \\
$\psi_{01}$ & $-0.0058$ ($-0.0168$, $0.0088$) & $0.2474$ ($0.1039$, $0.5098$) & $0.2558$ ($0.1631$, $0.5755$) & $0.9500$ ($0.9320$, $0.9610$) \\
$\psi_{02}$ & $0.0174$ ($-0.0028$, $0.0378$) & $1.2792$ ($0.8166$, $2.0289$) & $0.2842$ ($0.1823$, $0.4517$) & $0.9510$ ($0.9300$, $0.9630$) \\
$\psi_{11}$ & $-0.0052$ ($-0.0173$, $0.0078$) & $0.0636$ ($0.0100$, $0.1655$) & $0.1056$ ($0.0618$, $0.5676$) & $0.9450$ ($0.9300$, $0.9610$) \\
$\psi_{12}$ & $0.0174$ ($-0.0115$, $0.0589$) & $0.1885$ ($0.0762$, $0.5652$) & $0.3251$ ($0.1876$, $1.2615$) & $0.9485$ ($0.9370$, $0.9620$) \\
$\psi_{22}$ & $0.0418$ ($0.0199$, $0.0583$) & $1.2261$ ($0.7643$, $2.9078$) & $0.1432$ ($0.0876$, $0.3274$) & $0.9500$ ($0.9340$, $0.9620$) \\
$\theta_{\epsilon}$ & $-0.0006$ ($-0.0035$, $0.0024$) & $0.0490$ ($0.0224$, $0.1200$) & $0.0372$ ($0.0230$, $0.0600$) & $0.9490$ ($0.9370$, $0.9570$) \\
\hline
\hline
\multicolumn{5}{c}{\textbf{Latent Growth Curve Model}} \\
\hline
$\mu_{\eta_{0}}$ & $-0.0001$ ($-0.0003$, $0.0002$) & $0.2920$ ($0.2216$, $0.3685$) & $0.0058$ ($0.0044$, $0.0074$) & $0.9490$ ($0.9360$, $0.9590$) \\
$\mu_{\eta_{1}}$ & $-0.0006$ ($-0.0083$, $0.0058$) & $0.0624$ ($0.0245$, $0.1949$) & $0.0344$ ($0.0191$, $0.1906$) & $0.9475$ ($0.9350$, $0.9630$) \\
$\mu_{\eta_{2}}$ & $0.0002$ ($-0.0006$, $0.0026$) & $0.3153$ ($0.1786$, $1.0393$) & $-0.0105$ ($-0.0347$, $-0.0059$) & $0.9510$ ($0.9360$, $0.9630$) \\
$c$ & $0.0000$ ($-0.0010$, $0.0018$) & $0.0100$ ($0.0000$, $0.0265$) & $-0.0137$ ($-0.0384$, $-0.0070$) & $0.9485$ ($0.9310$, $0.9590$) \\
$\psi_{00}$ & $-0.0033$ ($-0.0097$, $0.0012$) & $2.1537$ ($1.5805$, $2.7190$) & $0.0864$ ($0.0632$, $0.1091$) & $0.9410$ ($0.9270$, $0.9600$) \\
$\psi_{01}$ & $-0.0063$ ($-0.0155$, $0.0070$) & $0.2475$ ($0.1039$, $0.5085$) & $0.2558$ ($0.1630$, $0.5731$) & $0.9495$ ($0.9330$, $0.9630$) \\
$\psi_{02}$ & $-0.0053$ ($-0.0176$, $0.0139$) & $1.2597$ ($0.7977$, $1.9744$) & $0.2799$ ($0.1772$, $0.4386$) & $0.9480$ ($0.9220$, $0.9630$) \\
$\psi_{11}$ & $-0.0053$ ($-0.0215$, $0.0064$) & $0.0636$ ($0.0141$, $0.1649$) & $0.1056$ ($0.0618$, $0.5632$) & $0.9445$ ($0.9300$, $0.9630$) \\
$\psi_{12}$ & $-0.0054$ ($-0.0409$, $0.0154$) & $0.1853$ ($0.0748$, $0.5498$) & $0.3200$ ($0.1840$, $1.2216$) & $0.9460$ ($0.9310$, $0.9570$) \\
$\psi_{22}$ & $-0.0036$ ($-0.0130$, $0.0037$) & $1.1681$ ($0.7294$, $2.7687$) & $0.1298$ ($0.0814$, $0.3075$) & $0.9445$ ($0.9320$, $0.9530$) \\
$\theta_{\epsilon}$ & $-0.0009$ ($-0.0045$, $0.0018$) & $0.0490$ ($0.0224$, $0.1200$) & $0.0372$ ($0.0230$, $0.0600$) & $0.9475$ ($0.9350$, $0.9600$) \\
\hline
\hline
\end{tabular}
\label{tbl:Metric_JB}
\begin{tablenotes}
\small
\item[a] SE: standard error.\\
\item[b] RMSE: root mean square error.\\
\end{tablenotes}
\end{threeparttable}}
\end{table}

\end{document}